\documentclass[
  aps,            %
  prmaterials,    %
  raggedbottom,   %
  reprint,        %
  superscriptaddress, %
  longbibliography, %
]{revtex4-2}

\usepackage{amsmath,amssymb}
\usepackage{graphicx}
\usepackage{xcolor}
\usepackage{booktabs}
\usepackage{tikz}
\usepackage{hyperref}

\definecolor{basepink}{RGB}{248,222,211}
\definecolor{basecyan}{RGB}{188,223,244}
\definecolor{gaadred}{RGB}{214,54,42}
\definecolor{gasym}{RGB}{232,176,155}
\definecolor{nsym}{RGB}{146,199,226}

\DeclareRobustCommand{\utri}{\raisebox{-0.02em}{\tikz{%
  \fill[gasym] (0,0) -- (0.66em,0) -- (0.33em,0.57em) -- cycle;
  \draw[black!75, line width=0.35pt]
    (0,0) -- (0.66em,0) -- (0.33em,0.57em) -- cycle;}}}
\DeclareRobustCommand{\dtri}{\raisebox{-0.02em}{\tikz{%
  \fill[nsym] (0,0.57em) -- (0.66em,0.57em) -- (0.33em,0) -- cycle;
  \draw[black!75, line width=0.35pt]
    (0,0.57em) -- (0.66em,0.57em) -- (0.33em,0) -- cycle;}}}

\DeclareRobustCommand{\tilA}{\raisebox{0.045em}{\tikz{%
  \fill[gray!25] (0,0.47em) -- (0.54em,0.47em) -- (0.81em,0)
    -- (0.27em,0) -- cycle;
  \draw[black!75, line width=0.45pt] (0,0.47em) -- (0.54em,0.47em)
    -- (0.81em,0) -- (0.27em,0) -- cycle;}}}
\DeclareRobustCommand{\tilB}{\raisebox{-0.19em}{\tikz{%
  \fill[gray!25] (0.27em,0.94em) -- (0.54em,0.47em) -- (0.27em,0)
    -- (0,0.47em) -- cycle;
  \draw[black!75, line width=0.45pt] (0.27em,0.94em)
    -- (0.54em,0.47em) -- (0.27em,0) -- (0,0.47em) -- cycle;}}}
\DeclareRobustCommand{\tilC}{\raisebox{0.045em}{\tikz{%
  \fill[gray!25] (0.27em,0.47em) -- (0.81em,0.47em) -- (0.54em,0)
    -- (0,0) -- cycle;
  \draw[black!75, line width=0.45pt] (0.27em,0.47em)
    -- (0.81em,0.47em) -- (0.54em,0) -- (0,0) -- cycle;}}}

\newcommand{\ntil}{n_{\mathrm{til}}}
\newcommand{\Iiso}{I_{\mathrm{iso}}}
\newcommand{\Gfull}{\mathcal{G}}
\newcommand{\nadj}{n_\mathrm{adj}}
\newcommand{\cfg}[1]{\texttt{#1}}
\newcommand{\Gprime}{\mathcal{G}'}
\newcommand{\Dthree}{D_3}
\newcommand{\Dsix}{D_6}
\newcommand{\Pfree}{\Pi_{\mathrm{free}}}
\newcommand{\nflex}{n_{\mathrm{flex}}}
\newcommand{\VHH}{V_{\mathrm{HH}}}
\newcommand{\dGaH}{\bar{d}_{\mathrm{GaH}}}

\begin{document}

\title{Tiling decomposition multiplicity predicts stability of
  GaN(0001) surface reconstructions}

\author{Tetsuji Kuboyama}
\email[Corresponding author: ]{kuboyama@tk.cc.gakushuin.ac.jp}
\affiliation{Computer Centre, Gakushuin University,
  1-5-1 Mejiro, Toshima-ku, Tokyo 171-8588, Japan}

\author{Akira Kusaba}
\affiliation{Research Institute for Applied Mechanics,
  Kyushu University, 6-1 Kasuga-koen, Kasuga,
  Fukuoka 816-8580, Japan}

\author{Karol Kawka}
\affiliation{Institute of High Pressure Physics,
  Polish Academy of Sciences, Sokolowska 29/37,
  01-142 Warsaw, Poland}

\author{Pawe\l{} Kempisty}
\affiliation{Institute of High Pressure Physics,
  Polish Academy of Sciences, Sokolowska 29/37,
  01-142 Warsaw, Poland}

\date{September 23, 2026}

\begin{abstract}
The electron counting (EC) rule helps explain semiconductor surface
stabilization and favored adsorbate coverages, but does not select
the preferred arrangement when many configurations satisfy it.
Motivated by the idea that satisfying the rule locally favors
stability, we enumerate configurations of GaN(0001)-$(6\times6)$
with 3~Ga adatoms and 18~H atoms that admit decomposition into
EC-balanced $(2\times2)$ blocks.  Even this restricted space
contains 416,683 symmetry-distinct configurations, making
energy-based searches costly.  We introduce the tiling decomposition
multiplicity, $\ntil$, which counts the compatible decompositions,
and evaluate it for every configuration.  Among 2,529 configurations
relaxed with a DFT-validated machine-learning interatomic potential,
the lowest energy in 13 of the 14 Ga placement symmetry classes is
attained by a configuration with maximal $\ntil$ within that class.  In the
remaining
class, a high-multiplicity configuration lies 8.5~meV below the
maximizer (17.5~meV in DFT), well under $k_B T$ at growth
temperature.  The energy ordering is largely associated with
avoiding adjacent bare surface Ga sites: each pair costs about
0.2~eV.  The existence of a compatible decomposition is a separate
global requirement.  The 24
multiplicity maximizers are exhaustively identified and all
evaluated with DFT.  Counting compatible decompositions thus narrows
416,683 configurations to 24 promising candidates for
first-principles evaluation, without computing the energy of every
configuration.
\end{abstract}

\maketitle

\section{Introduction}
\label{sec:intro}

Surface reconstructions of III-nitride semiconductors critically
influence the growth, doping, and device performance of GaN-based
optoelectronic and power electronic
devices~\cite{Zywietz1998,VanDeWalleJVST2002,Neugebauer2003,%
Akiyama2024CGD,Kangawa2025}.  On
the Ga-terminated GaN(0001) surface under metalorganic vapor phase
epitaxy (MOVPE) conditions, Ga adatoms and H adsorbates
coexist~\cite{Kusaba2022,Kawka2024}, and
surface stabilization is governed by the electron counting (EC)
rule~\cite{Pashley1989,Zhang2006GEC,Kangawa2013}: adsorbate-mediated charge
compensation leaves the dangling bonds of the topmost Ga atoms
unoccupied, yielding a stable semiconducting surface.  The EC rule
constrains the surface coverage and has been used to examine
candidates for stable reconstructions, primarily within the
$(2\times 2)$ surface~\cite{Kusaba2017}.  The rule does not,
however, select the preferred arrangement when
many configurations satisfy it.  We ask what structural property
distinguishes the most stable configurations.

On the $(6\times 6)$ surface with 3~Ga adatoms and 18~H atoms,
Bayesian-optimization searches found that stable configurations
favor local satisfaction of the EC rule~\cite{Kusaba2022}.
Kawka~\textit{et~al.}~\cite{Kawka2024} assessed this by decomposing
the surface into EC-balanced $(2\times 2)$ tiles and found that
inter-tile interactions also affect stability.  Even restricting
the search to configurations admitting such decompositions leaves
hundreds of thousands of possibilities, with computed relative
energies spanning about 1~eV per cell.  Sampling cannot guarantee
reaching the most stable configuration within a limited number of
evaluations.  Our objective is a descriptor that ranks promising
configurations from the adatom positions alone, together with a
physical explanation of the ranking.

One configuration can admit several EC-balanced decompositions,
a fact first noted by Kawka~\textit{et~al.} under the name
\emph{number of arrangements}~\cite{Kawka2024}.  We call their
number the \emph{tiling decomposition multiplicity} $\ntil$.
We compare H arrangements within classes that share the same
Ga-adatom placement up to surface symmetry.
Enumerating all 416,683 configurations that admit such a
decomposition (EC-compatible configurations, Sec.~\ref{sec:tiling})
and relaxing
2,529 selected configurations with a validated machine-learning
interatomic potential (MLIP), we find that in 13 of the 14 symmetry
classes of Ga adatom placements the lowest energy among the relaxed
configurations is attained by a configuration of maximal $\ntil$.
In the fourteenth, the minimum is among the highest-$\ntil$
configurations and sits 8.5~meV below the maximizer.

This ordering is largely associated with the avoidance of
nearest-neighbor pairs of \emph{bare sites}, topmost Ga atoms that
carry no H atom and adjoin no Ga adatom.  Each pair costs about
0.2~eV, every class minimum is free of such pairs, and $\ntil$ tends
to be large when the bare sites stay apart.  Whether a compatible
decomposition exists at all is a separate, global condition, worth
an energy of order 0.1~eV near the minima.

Exhaustive enumeration supplies all maximizers of $\ntil$:
24 candidates across the 14 classes, all of which we evaluated with
density functional theory (DFT).  This completeness concerns the
catalog and the descriptor; the energetic rule is tested on the
relaxed subset.  Symmetry-reduced exhaustive enumeration is well
established for bulk configuration spaces, for instance in the
derivative-structure enumeration of alloys~\cite{HartForcade2008};
to our knowledge it has not been applied before to the
reconstruction problem of a semiconductor surface under the EC
constraint.

Counting the ways in which a global constraint can be met locally
has a precedent in mathematical chemistry, where the Kekul\'{e}
structure count of benzenoid hydrocarbons correlates with
thermodynamic stability within classes of
isomers~\cite{Herndon1973,Swinborne1975,Randic2003}.  In a separate
line of work, Mora-Fonz~\textit{et~al.}~\cite{MoraFonz2017} showed
that configurational entropy, arising from the many
nonstoichiometric configurations that satisfy the
dipole-cancellation constraint, stabilizes the polar surfaces of
ZnO.  Our $\ntil$ is neither of these: it counts decompositions of
one and the same physical state, not distinct states or quantum
superpositions.  We return to both parallels in
Secs.~\ref{sec:kekule} and~\ref{sec:entropy}.

Section~\ref{sec:methods} defines the lattice model, the
enumeration, and the energy evaluation; Sec.~\ref{sec:results}
presents the catalog, the validation, and the $\ntil$-max rule
with its local mechanism; Sec.~\ref{sec:discussion} discusses the
physical picture and the practical implications.

\section{Model and Methods}
\label{sec:methods}

We define the lattice model and $\ntil$, construct the catalog,
and describe the MLIP and DFT calculations.  Readers interested
chiefly in the physical results may defer the enumeration details
of Sec.~\ref{sec:enumeration} until after Sec.~\ref{sec:results}.

\subsection{Lattice model and the EC rule}
\label{sec:lattice}

The GaN(0001)-$(6\times 6)$ surface is modeled as a triangular
lattice on a torus, that is, with periodic boundary conditions in
both lattice directions, spanning $6 \times 6$ surface cells and
comprising 72
triangular faces: 36 upward-pointing (\utri, Ga sublattice) and 36
downward-pointing (\dtri, N sublattice).  Each \utri\ face hosts a
topmost Ga atom of the crystal at its center, and each \dtri\ face
an N atom of the layer below; the correspondence between the atomic
arrangement and the face lattice is shown in
Fig.~\ref{fig:lattice}(a).  The face indexing used in the catalog
is given in Appendix~\ref{app:symmetry}.

\begin{figure*}[tb]
\includegraphics[width=\textwidth]{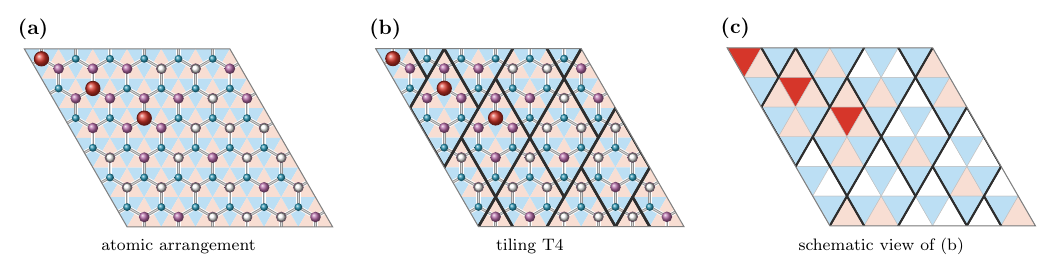}
\caption{\label{fig:lattice}
  From the atomic arrangement to the puzzle model.
  (a)~Atomic arrangement of an EC-compatible (Sec.~\ref{sec:tiling})
  configuration $(G, H)$
  on the GaN(0001)-$(6 \times 6)$ surface (configuration \cfg{109117},
  the configuration with the lowest MLIP energy found;
  Sec.~\ref{sec:ntil_rule}).
  Magenta spheres are the topmost Ga atoms, located at the centers
  of the upward-pointing faces \utri\ (light red, Ga sublattice);
  teal spheres are the N atoms of the layer below, at the centers of
  the downward-pointing faces \dtri\ (light blue, N sublattice).
  Large
  dark-red spheres are the 3~Ga adatoms, each above a \dtri\ face;
  small white spheres are the 18~H atoms, on top of the Ga atoms of
  the designated \utri\ faces.  Periodic boundary conditions
  identify opposite edges of the parallelogram.
  (b)~The same configuration decomposed into nine blocks by one of
  its compatible tilings (tiling class T4;
  Table~\ref{tab:tiling_classes} in Appendix~\ref{app:symmetry}),
  with block boundaries drawn as
  black lines.
  (c)~Schematic representation of (b), used in the remainder of
  this paper: a dark red \dtri\ face hosts a Ga adatom, a white
  \utri\ face hosts an H atom, and bare faces retain the pale
  sublattice colors.}
\end{figure*}

The $(6 \times 6)$ cell is not proposed as an observed
reconstruction; it is the computational cell adopted in the
preceding studies~\cite{Kusaba2022,Kawka2024}, large enough to host
a mixture of $(2 \times 2)$-type local units and hence the smallest
scale on which the decomposition multiplicity studied below becomes
nontrivial.  Under MOVPE conditions, Ga and H co-adsorb on the
surface; here we focus on a single stoichiometry that satisfies the
EC rule on the $(6 \times 6)$ cell, with each configuration
consisting of 3~Ga adatoms occupying \dtri\ faces and 18~H atoms
occupying \utri\ faces.  This composition corresponds to a mixture
of three Ga$_\mathrm{ad}$-$(2 \times 2)$-type~\cite{Rapcewicz1997,Feenstra1998}
and six
3Ga-H-$(2 \times 2)$-type~\cite{Fritsch1998} units, the two reconstructions
whose free
energies lie close to each other near their phase boundary under
MOVPE conditions~\cite{Kusaba2017,Kusaba2022,Kawka2024}.  A Ga adatom at a
\dtri\ face bonds to the three
adjacent \utri\ faces (the underlying Ga atoms of the top bilayer),
while H atoms passivate the dangling bonds of the \utri\ faces not
adjacent to any Ga adatom.

The EC-compatible stoichiometries follow from electron bookkeeping.
Each surface Ga dangling bond holds $3/4$ of an electron, which the
EC rule requires to be transferred away; a bare \utri\ face
therefore releases $3/4$ of an electron, whereas a Ga--H bond
absorbs $1/4$ and a Ga-adatom unit absorbs $3/4$.  With
$n_\mathrm{Ga}$ adatoms and $n_\mathrm{H}$ hydrogens on the 36
\utri\ faces, the balance
$(36 - 3 n_\mathrm{Ga} - n_\mathrm{H}) \cdot 3/4 =
n_\mathrm{H}/4 + 3 n_\mathrm{Ga}/4$ reduces to
\begin{equation}
  3\,n_\mathrm{Ga} + n_\mathrm{H} = 27,
  \label{eq:ec_line}
\end{equation}
so the number of bare \utri\ faces,
$36 - 3 n_\mathrm{Ga} - n_\mathrm{H} = 9$, is the same for every
EC-compatible composition.  The composition studied here is the
point $n_\mathrm{Ga} = 3$, $n_\mathrm{H} = 18$ on this line
(other compositions correspond to $k_G = n_\mathrm{Ga} = 0$--$9$;
Sec.~\ref{sec:outlook}).

Two configurations that differ only by a symmetry operation of the
lattice describe the same surface and are counted once.  The
operations that matter here are the 36 translations of the torus,
the rotations by multiples of $120^\circ$, and the three reflections
that carry each sublattice into itself.  The $60^\circ$ rotation and
the remaining reflections exchange the two sublattices; because Ga
adatoms occupy \dtri\ faces and H atoms \utri\ faces, these do not
relate physically equivalent configurations and are excluded.  The
216 operations that remain form the group $\Gprime$, used throughout
to identify equivalent configurations.  Its structure, its action on
a tiled configuration, and the counting identities against which the
enumeration was checked are given in Appendix~\ref{app:symmetry}.

\subsection{Rhombus tilings and the multiplicity $\ntil$}
\label{sec:tiling}

The decompositions into locally EC-satisfying $(2 \times 2)$ blocks
described in Sec.~\ref{sec:intro} are made precise here as rhombus
tilings of the face lattice.

\begin{figure}[tb]
\includegraphics[width=0.86\columnwidth]{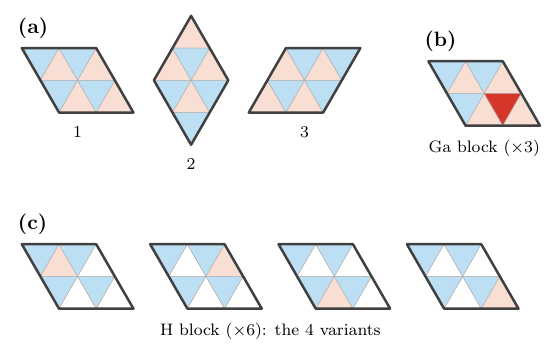}
\caption{\label{fig:pieces}
  The pieces of the rhombus-tiling model, in the schematic
  representation of Fig.~\ref{fig:lattice}(c).  Each rhombus block
  covers eight faces ($4\,\utri + 4\,\dtri$) and contains exactly
  one \dtri\ face surrounded by three \utri\ faces, the Ga-adatom
  site.
  (a)~The three piece orientations $1$~(\tilA), $2$~(\tilB), and
  $3$~(\tilC), related by $120^\circ$ rotations.
  (b)~A Ga block hosts a Ga adatom at that site (dark red
  \dtri\ face) and carries no H.
  (c)~An H block hosts three H atoms (white \utri\ faces) on three
  of its four \utri\ faces; the choice of the bare \utri\ face
  (pale) gives the four variants shown.  A tiling of the torus uses
  9~blocks: 3~Ga blocks and 6~H blocks.}
\end{figure}

A \emph{rhombus tiling} of the $(6 \times 6)$ torus is a partition of
all 72 faces into 9 non-overlapping rhombus blocks, each consisting of
4~\utri\ faces and 4~\dtri\ faces.  Each rhombus takes one of
three orientations, labeled $1$~(\tilA), $2$~(\tilB), and
$3$~(\tilC) [Fig.~\ref{fig:pieces}(a)], and within each block one
\dtri\ face (surrounded by three \utri\ faces) serves as the
potential Ga-adatom site.  The local EC rule for a given tiling
requires that exactly 3 of the 9 blocks host Ga adatoms, with the
remaining 6 blocks hosting 3~H atoms each on their \utri\ faces;
Figs.~\ref{fig:pieces}(b) and \ref{fig:pieces}(c) show the two
resulting block types.  A
tiling that satisfies the local EC rule at every block therefore
determines an adatom configuration $(G, H)$, up to the $4^6$ choices
of which \utri\ face in each H block remains bare.
Figures~\ref{fig:lattice}(b) and \ref{fig:lattice}(c) show the
configuration of Fig.~\ref{fig:lattice}(a) decomposed into nine
blocks by one of its compatible tilings, in the atomistic and the
schematic representation, respectively.  The restriction to this
single block shape is a modeling choice, anchored to the two
$(2 \times 2)$ reconstructions underlying the composition
(Sec.~\ref{sec:lattice}); other EC-satisfying local units are not
considered.

\begin{figure*}[t!]
\includegraphics[width=\textwidth]{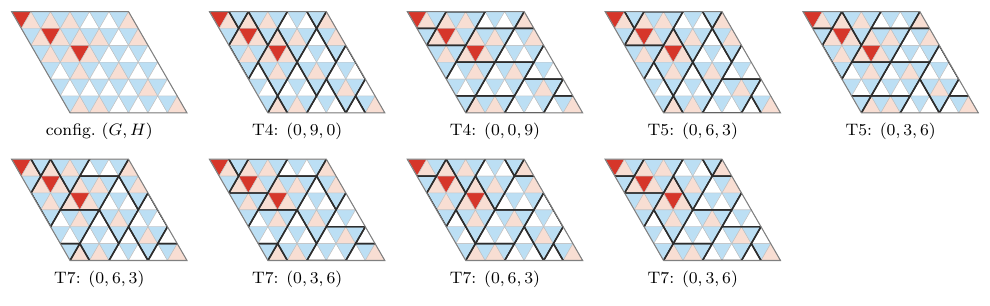}
\caption{\label{fig:interpretations}
  All eight rhombus tilings compatible with a single EC-compatible
  configuration $(G, H)$ (top left; configuration \cfg{109117},
  the configuration with the lowest MLIP energy found,
  $\ntil = 8$).  The Ga and H positions are
  identical in every panel; only the partition into rhombus blocks
  differs.  Each tiling is labeled by its tiling class
  (Table~\ref{tab:tiling_classes} in Appendix~\ref{app:symmetry})
  and orientation distribution
  $(n_1, n_2, n_3)$, which records how many of the 9 blocks take
  each of the three orientations of Fig.~\ref{fig:pieces}(a).  The tiling
boundaries
  are not physical objects: they are distinct decompositions of the
  global EC constraint into block-local balance equations, and
  $\ntil$ counts these decompositions for a fixed physical state.}
\end{figure*}

The converse assignment is the origin of the multiplicity studied in
this work.  Given a configuration $(G, H)$, a tiling $\mathcal{T}$ is
\emph{compatible} with $(G, H)$ if the block contents dictated by the
atom positions satisfy the local EC rule in every block: each block
either hosts one Ga adatom at its Ga-adatom site and no H (a Ga
block), or hosts exactly three H atoms on its \utri\ faces (an H
block), with exactly 3 Ga blocks in total.  A configuration is
\emph{EC-compatible} if at least one compatible tiling exists; the
term refers to this local decomposability, since the global electron
count is already fixed by the stoichiometry [Eq.~(\ref{eq:ec_line})].  The
map from tilings to configurations is many-to-one; the same physical
configuration can be reproduced from several distinct
tilings.  We call the number of compatible tilings the \emph{tiling
decomposition multiplicity},
\begin{equation}
  \ntil(G, H) = \bigl|\{\mathcal{T} :
    \mathcal{T} \text{ is compatible with } (G, H)\}\bigr|,
  \label{eq:ntil}
\end{equation}
which takes integer values from 1 to 66 for configurations on the
$(6 \times 6)$ torus.

Figure~\ref{fig:interpretations} shows all eight compatible tilings
of the configuration in Fig.~\ref{fig:lattice}(a).  The atom
positions are unchanged: the tilings represent different
decompositions of one physical state into locally balanced units,
not additional physical degrees of freedom.

\subsection{Exhaustive enumeration}
\label{sec:enumeration}

\begin{figure}[tb]
\includegraphics[width=0.94\columnwidth]{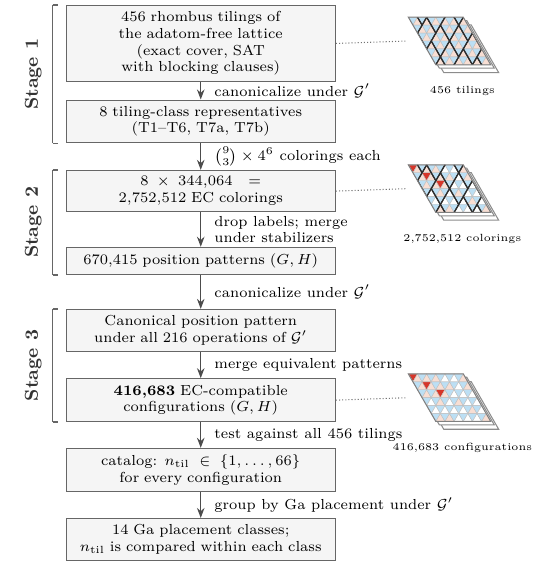}
\caption{\label{fig:pipeline}
  The enumeration procedure (Sec.~\ref{sec:enumeration}).  Boxes
  give the objects and their counts, arrow labels name the
  operations, side rails mark the three stages, and the thumbnails
  show an example object of the corresponding stage [the
  configuration of Fig.~\ref{fig:lattice}(a) and one of its
  tilings]: a tiling of the adatom-free lattice, a coloring (tiling + adatoms),
  and the configuration obtained by dropping the tiling label.
  The final step groups the catalog by Ga placement into the 14
  classes ($\Gprime$-orbits), within which the multiplicity
  $\ntil$ is compared (Sec.~\ref{sec:ntil_rule}).}
\end{figure}

The enumeration runs in three stages
(Fig.~\ref{fig:pipeline}): the tilings of the adatom-free lattice
(the lattice before any adatom is placed), the EC-compatible adatom
configurations built on them, and the reduction of the result to
symmetry-distinct configurations.  The technical steps are collected
in Appendix~\ref{app:symmetry}.

\begin{figure*}[tb]
\includegraphics[width=\textwidth]{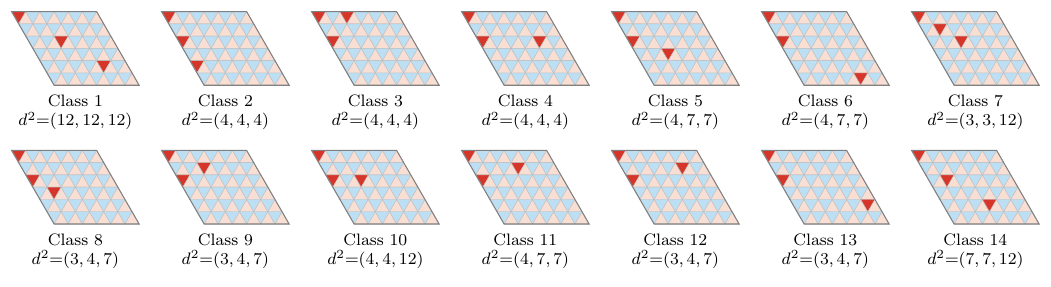}
\caption{\label{fig:ga_classes}
  Canonical representatives of the 14 Ga placement classes
  (Table~\ref{tab:ga_orbits}): the three Ga adatom positions (dark
  red \dtri\ faces) are shown; H arrangements are not part of
  the class definition.  Each class is the $\Gprime$-orbit of a
  3-element subset of the 36 \dtri\ faces, labeled by the
  squared pair-distance signature $d^2$ of the three Ga adatoms.  Classes
  sharing the same $d^2$ (e.g., Classes~2--4) are inequivalent
  arrangements with identical pair distances.}
\end{figure*}

\emph{Stage~1: Tilings of the adatom-free lattice.}---The rhombus
tilings of the empty $(6 \times 6)$ torus are enumerated as an exact
cover: each of the 72 faces has to belong to exactly one of the
$36 \times 3 = 108$ candidate rhombus pieces, and no face may be
covered twice.  We solve this with a Boolean satisfiability (SAT)
solver, a general-purpose program that takes a set of logical
constraints on true/false variables, decides whether they can all be
met at once, and returns an assignment of the variables when they
can.  Each candidate piece is one variable here, true when the piece
is used, and the exact-cover condition is written as logical clauses
over those variables.  A solver returns one solution rather than all
of them, so after each solution we add a clause that rules out
precisely that assignment---a blocking clause---and run the solver
again, until it reports that none is left; the number of solutions
returned along the way is the number of tilings.  We used the open-source
solver Glucose
4.2.1~\cite{Glucose} (cf.\ the per-structure SAT decomposition of
Ref.~\cite{Kuboyama2024}).  The tilings of the adatom-free lattice
are enumerated once and for all, independently of the adatoms, and
there are exactly 456 of them.  Under $\Gprime$ they fall into 8
classes (Appendix~\ref{app:symmetry}), and the next stage needs only
one representative of each, since every tiling is the
$\Gprime$-image of a representative.

\emph{Stage~2: Adatom configurations on the representative
tilings.}---For each of the 8 representatives we enumerate every
assignment of 3~Ga blocks and 6~H blocks allowed by the local EC
rule; a tiling carrying such an assignment is called a
\emph{coloring}.  There are
$\tbinom{9}{3} \times 4^6 = 344{,}064$ colorings per tiling [the
choice of the 3 Ga blocks and, in each H block, the choice of the
bare \utri\ face; Fig.~\ref{fig:pieces}(c)], 2,752,512 in all.
Each coloring is recorded as an adatom position pattern $(G,H)$,
without a tiling label, and reduced under the symmetries that
preserve its representative tiling.  Identical position patterns
within and across representatives are merged, leaving 670,415
patterns.

\emph{Stage~3: Symmetry reduction of position patterns.}---The
670,415 patterns are reduced to one representative per
$\Gprime$-orbit, giving the final catalog of 416,683 physically
distinct EC-compatible configurations.  Each of them carries a
unique identifier, cited in this paper in typewriter type (e.g.,
configuration \cfg{109117}) to keep identifiers apart from numerical
values; the catalog, with the adatom coordinates and the $\ntil$
values, is openly available (see the Data availability statement).
The multiplicity of each configuration is then obtained by testing
it for compatibility against all 456 tilings; the resulting
distribution of $\ntil$ is analyzed in
Sec.~\ref{sec:config_space}.

We classify Ga adatom placements into equivalence classes under
$\Gprime$: two placements $G_1$ and $G_2$ belong to the same
\emph{Ga placement class} if $g(G_1) = G_2$ for some
$g \in \Gprime$.  Each class is an \emph{orbit} of $\Gprime$ acting
on 3-element subsets of the 36~\dtri\ faces, and its size $|O|$ is
the number of symmetry-equivalent placements it contains;
``orbit'' is used in this group-theoretic sense throughout and is
unrelated to electronic orbitals.  These placement classes partition
the adatom configurations by their Ga geometry, and are distinct
from the tiling classes of Stage~1, which partition the 456 tilings.
For
the present composition of $k_G = 3$ Ga adatoms, this action yields
47 such orbits, of which exactly 14
admit at least one EC-compatible $H$ configuration.  We label these
14 as Class~1--14, in order of increasing orbit size $|O|$ and of
the lexicographically smallest placement in the class, which we take
as its canonical representative; together they account for all
416,683
EC-compatible adatom configurations.  Each class is characterized by
the multiset of squared pair distances of its three Ga adatoms, the
signature $d^2$ (distances measured in units of the surface
lattice constant);
canonical representatives are shown in Fig.~\ref{fig:ga_classes}
and the class properties are summarized in
Table~\ref{tab:ga_orbits}.

All counts in this section are exact combinatorial statements
involving no energy model, and were checked independently as
described in Appendix~\ref{app:symmetry}.

The enumeration is inexpensive next to the MLIP energy evaluation
of Sec.~\ref{sec:mlip}.  On a laptop with an Apple M4 Max (16 CPU cores), the
whole pipeline, from the empty lattice to the catalog with $\ntil$
assigned, runs in about five minutes.  The exact-cover enumeration
of the 456 tilings takes under 0.01~s; Stage~2, the 2,752,512
colorings and their reduction to 670,415 position
patterns, 25~s on one core; Stage~3, the reduction under
$\Gprime$ to the 416,683 configurations, 13~s on 16 cores; the
index that records, for each configuration, the representative
tilings it derives from, 200~s on one core; and the assignment of
$\ntil$ to every configuration, by testing it against all 456
tilings, 1~minute on 16 cores.

\subsection{MLIP structural relaxation}
\label{sec:mlip}

The atomic structure corresponding to each configuration $(G, H)$ was
constructed as a six-bilayer GaN(0001) slab with $(6 \times 6)$
lateral periodicity and a vacuum layer of 20~\AA.  Ga adatoms were
placed at T$_4$ sites (the threefold hollow site above a
second-layer N atom) over the designated \dtri\ faces, and H atoms
at on-top sites, directly above the surface Ga atoms of the
designated \utri\ faces.

The slab exposes the Ga-polar $(0001)$ face at the top and the
N-polar $(000\bar{1})$ face at the bottom.  The N-polar face was
modeled as the 3N-H $(2 \times 2)$ reconstruction, in which H atoms
occupy three of every four N sites.  This reconstruction satisfies
the EC rule on the N-polar face, so that the face is semiconducting
and stable~\cite{Northrup2004}.  Slab DFT usually terminates the
bottom face with pseudo-hydrogen atoms, which carry a noninteger
nuclear charge chosen to match the dangling bonds exactly; the MLIP
does not support pseudo-atoms, and a physical reconstruction was
used instead.  The central two bilayers of the slab were fixed at
bulk positions, and every other atom, including the adsorbates on
both faces, was allowed to relax, so that the bottom face relaxes to
its own reconstructed state rather than being held in a strained
one.  Because every configuration carries the same bottom face, its
contribution is expected largely to cancel in the energy differences
between configurations.

Structural relaxation was performed with the BFGS optimizer, a
quasi-Newton method, as implemented in the Atomic Simulation
Environment (ASE)~\cite{ASE2017},
using the UMA (Universal Models for Atoms) machine-learning
interatomic potential~\cite{UMA2024} (model \texttt{uma-m-1.1} with
the \texttt{oc20} task, which targets adsorbate--surface systems)
for energy and force evaluation, with a force convergence criterion
of $f_\mathrm{max} = 0.05$~eV/\AA.  The reliability of this
potential for the present chemistry is established by the direct
DFT validation of Sec.~\ref{sec:dft}, not assumed from its training
domain.

A total of 2,529 configurations were selected for MLIP relaxation
by a stratified procedure designed to cover the full range of $\ntil$
values within each class.  Within each of the 14 Ga placement
classes, configurations were ranked by a composite score that
combines $\ntil$ with the three geometric descriptors defined in
Sec.~\ref{sec:isotropy} ($\Iiso$, $\VHH$, and $\dGaH$), with
weights estimated from a preliminary regression.  The top-ranked configurations
were
selected first.  To avoid a systematic bias toward high-$\ntil$
configurations, we additionally included the configurations with the
lowest $\ntil$ within each class and a stratified random sample
across intermediate $\ntil$ values.  The final sample spans the
complete $\ntil$ range in every class, from $\ntil = 1$ to the
class-specific maximum, and contains all 24 configurations that
attain the class maximum of $\ntil$ in their respective classes
(Sec.~\ref{sec:practical}).  All 2,529 calculations reached the
force convergence criterion.  The additional targeted relaxations
reported in Sec.~\ref{sec:nadj} (the 76 members of the complete
$\nadj = 0$ set not already contained in the sample, and the
untileable arrangements, that is, adatom patterns admitting no
compatible tiling and therefore not EC-compatible) used the
identical protocol.

No interatomic potential was trained in this work.  The released
\texttt{uma-m-1.1} model is used as it stands, and the only data
preparation is the stratified selection described above.
On a single NVIDIA GeForce RTX 3090, with one configuration relaxed
at a time, each MLIP relaxation of the 480-atom slab took 343~s and
156 BFGS steps on average; the 2,529 relaxations therefore came to about
240 GPU hours.  The recorded runtimes for the 164 additional
configurations reported in Sec.~\ref{sec:nadj} total about 13 GPU hours.

\subsection{DFT reference data}
\label{sec:dft}

We validate the MLIP against two independent DFT data sets,
710 structures in all.  The fixed-placement set tests relative
energies across many H arrangements at one Ga placement; the
cross-class set tests the 24 multiplicity maximizers and the
Class-2 exception.  For the $\ntil$-max rule, the essential test
is whether the MLIP preserves the relative energy ordering within
each class, particularly among low-energy configurations.

\emph{The fixed-placement set: 685 structures at one Ga
placement.}---The first set is a database of 685 DFT-computed adatom structures
on the
$(6 \times 6)$ surface in which the three Ga adatoms are fixed at a
Class-7 placement [the diagonal $d^2 = (3,3,12)$ arrangement of the
MLIP global minimum, Fig.~\ref{fig:lattice}(a)] and the 18~H atoms take
685 arrangements.  The structures and their DFT energies
are taken from the two Bayesian-optimization sampling runs of
Kawka~\textit{et~al.}~\cite{Kawka2024}, which explored H
arrangements at this fixed Ga placement.
The DFT energies were computed with the real-space DFT code
RSDFT~\cite{Iwata2010} within the generalized gradient
approximation, using the PBE functional~\cite{PBE1996} and
norm-conserving pseudopotentials, on a GaN(0001)-$(6 \times 6)$ slab
with $\Gamma$-point sampling and a real-space mesh spacing of
0.15~\AA\ (equivalent to a 128~Ry cutoff); full computational
details are given in Refs.~\cite{Kusaba2022,Kawka2024}.
The set spans 2.5~eV in relative energy and is not restricted to the
EC-compatible space; only 83 of the 685 structures admit a
compatible tiling (81 distinct configurations, two of them sampled
twice), while the remaining 602 are EC-incompatible.
This set therefore tests the MLIP over chemically diverse local
environments, and it simultaneously probes, at the
DFT level, the energetic separation between EC-compatible and
EC-incompatible arrangements (Sec.~\ref{sec:ntil_rule}).  All 685
structures were re-evaluated with the same slab model, MLIP, and
relaxation protocol as in Sec.~\ref{sec:mlip}, except for one
constraint inherited from the source database: the in-plane
coordinates of the three Ga adatoms were held fixed during the MLIP
relaxation, as they were in the DFT calculations of
Ref.~\cite{Kawka2024}, in which the placement of the Ga adatoms was
treated as a discrete variable of the search.
Matching this constraint permits a structure-by-structure comparison
with the reference DFT energies.  The
restriction was used only for this MLIP validation set; the adatom
positions were free in the 2,529 MLIP relaxations of
Sec.~\ref{sec:mlip}, including the cross-class candidates.
The convergence criterion was unchanged
($f_\mathrm{max} = 0.05$~eV/\AA).

\emph{The cross-class set: 25 configurations across all 14
classes.}---The second set comprises 25 EC-compatible configurations
spanning all 14 Ga placement classes over a 655~meV range, computed
independently of the fixed-placement set.
The set consists of the complete family of 24 $\ntil$-max candidate
configurations (between one and four per class,
Table~\ref{tab:ga_orbits}; Sec.~\ref{sec:practical}) together with
the MLIP-predicted class minimum of Class~2, so that it covers both
configurations involved in the only exception to the $\ntil$-max
rule (Sec.~\ref{sec:ntil_rule}).  In the four classes whose class
maximum is $\ntil^\mathrm{max} = 16$ (Classes~5, 6, 9, and~12), the
candidates sharing this common $\ntil$ range from the MLIP-predicted
class minimum to strongly destabilized configurations; their
internal ordering therefore probes the MLIP fidelity independently
of the $\ntil$--energy correlation under test.

These DFT energies were computed with the SIESTA
code~\cite{Siesta2002} within the generalized gradient
approximation, using the PBEJsJrLO parameterization of the
exchange--correlation functional, which combines the
Perdew--Burke--Ernzerhof functional~\cite{PBE1996} with the
jellium-surface, jellium-response, and Lieb--Oxford
criteria~\cite{Pedroza2009,Odashima2009}.
Norm-conserving pseudopotentials were generated with the
valence configurations $2s^2\,2p^3$ for nitrogen and
$3d^{10}\,4s^2\,4p^1$ for gallium.  Triple-zeta basis sets were used
for the $s$ and $p$ orbitals of all elements and double-zeta sets
for the $d$ orbitals, and the real-space grid used a mesh cutoff of
410~Ry (a spatial resolution of about 0.09~\AA); the lattice
constants obtained with this parametrization were $a = 3.214$~\AA\
and $c = 5.233$~\AA.  The slab was a $(6 \times 6)$ GaN supercell of
three Ga--N bilayers whose N-terminated bottom surface was saturated
with 36 pseudo-hydrogen atoms of effective charge $0.75e$, for 252
atoms in the substrate and 273 atoms in total including the 3~Ga and
18~H surface adatoms.  The Brillouin zone was sampled at the
$\Gamma$ point.  During structural relaxation, only the bottom Ga--N
bilayer and pseudo-hydrogen atoms were fixed; all other atoms were
allowed to relax in all three directions.  Self-consistent-field (SCF)
tolerances were $10^{-4}$ for
the density matrix and $5 \times 10^{-4}$~eV for the Hamiltonian;
the force tolerance was $10^{-2}$~eV/\AA.
These calculations were
run on a single 28-core node of the Topola cluster at ICM,
University of Warsaw: one SCF iteration took 20--30~s, and one
DFT relaxation of the 273-atom slab about 6 hours of wall time, or
roughly 170 core hours.

The DFT energies are presented as mixing enthalpies of the
Ga-adatom~$(2 \times 2)$ and 3Ga-H~$(2 \times 2)$
surfaces~\cite{Kawka2024}; since the surface composition is fixed
here, they are simply relative energies.  Because the mixing
enthalpy of the most stable configuration in the DFT data set is
nearly zero, the MLIP energies are presented as relative energies
referenced to that most stable configuration.  In short,
$\Delta E_\mathrm{MLIP}$ and $\Delta E_\mathrm{DFT}$ can thus be
regarded as relative energies measured from the same reference
configuration.  The comparison
metrics are the Pearson (linear) and Spearman (rank) correlations,
the mean absolute error (MAE) of relative energies, and the
preservation of within-class energy ordering.
The quantitative results are presented in
Sec.~\ref{sec:mlip_dft}.

\section{Results}
\label{sec:results}

\subsection{The catalog of EC-compatible configurations}
\label{sec:config_space}

\begin{table}[tb]
\caption{\label{tab:ga_orbits}%
The 14 Ga placement classes.  For each class, the squared pair-distance
signature $d^2$ of the three Ga adatoms, the class size $|O|$ (number of
symmetry-equivalent Ga placements), the number of EC-compatible
configurations $N_\mathrm{pat}$, the class maximum
$\ntil^\mathrm{max}$ of the multiplicity, and the number
$N_{\ntil\mathrm{-max}}$ of configurations attaining that maximum
are listed.  The minimum of $\ntil$ is 1 in every class.  Repeated
values of $N_\mathrm{pat}$
across classes are exact coincidences: the counts were obtained
independently for each class and sum to 416,683.}
\begin{ruledtabular}
\begin{tabular}{lcrrrr}
Class & $d^2$ & $|O|$ & $N_\mathrm{pat}$ & $\ntil^\mathrm{max}$ &
$N_{\ntil\mathrm{-max}}$ \\
\hline
1 & $(12,12,12)$ & 12 & 6,691 & 42 & 1 \\
2 & $(4,4,4)$ & 36 & 17,332 & 66 & 1 \\
3 & $(4,4,4)$ & 36 & 20,381 & 66 & 1 \\
4 & $(4,4,4)$ & 36 & 19,548 & 66 & 1 \\
5 & $(4,7,7)$ & 108 & 16,061 & 16 & 3 \\
6 & $(4,7,7)$ & 108 & 16,061 & 16 & 3 \\
7 & $(3,3,12)$ & 108 & 14,896 & 8 & 1 \\
8 & $(3,4,7)$ & 216 & 29,791 & 8 & 1 \\
9 & $(3,4,7)$ & 216 & 32,116 & 16 & 4 \\
10 & $(4,4,12)$ & 216 & 122,317 & 58 & 1 \\
11 & $(4,7,7)$ & 216 & 29,791 & 8 & 1 \\
12 & $(3,4,7)$ & 216 & 32,116 & 16 & 4 \\
13 & $(3,4,7)$ & 216 & 29,791 & 8 & 1 \\
14 & $(7,7,12)$ & 216 & 29,791 & 8 & 1 \\
\end{tabular}
\end{ruledtabular}
\end{table}

\begin{figure}[tb]
\includegraphics[width=\columnwidth]{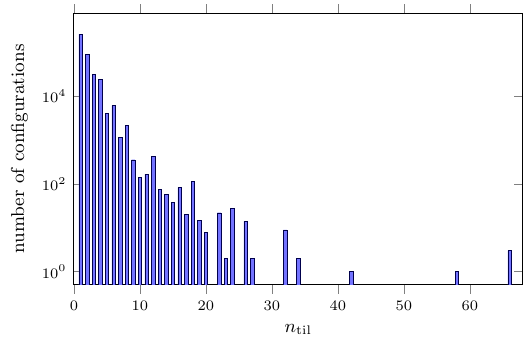}
\caption{\label{fig:ntil_hist}
  Distribution of the tiling decomposition multiplicity $\ntil$ over
  all 416,683 EC-compatible configurations (logarithmic vertical
  axis).  The majority of configurations (61.6\%) admit a unique
  decomposition ($\ntil = 1$); the maximum $\ntil = 66$ is attained
  by three configurations.}
\end{figure}

\begin{figure*}[tb]
\includegraphics[width=0.48\textwidth]{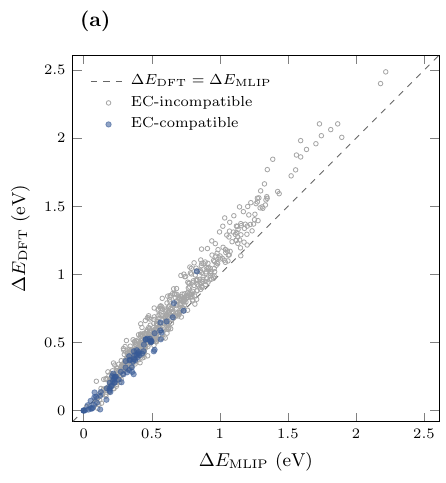}%
\hfill
\includegraphics[width=0.48\textwidth]{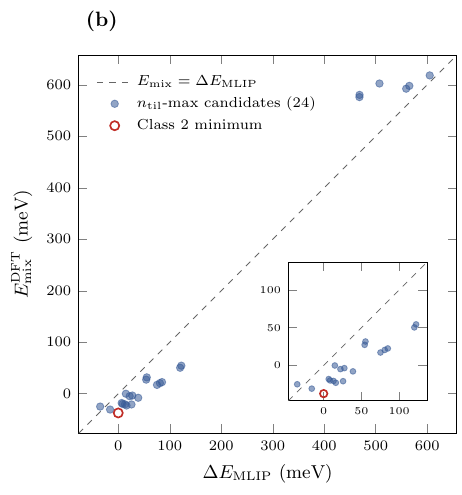}
\caption{\label{fig:mlip_dft}
  MLIP versus DFT relative energies for the two validation sets of
  Sec.~\ref{sec:dft}.  The dashed lines indicate perfect agreement.
  (a)~The fixed-placement set: 685 structures at a fixed Class-7
  Ga placement,
  including 602 EC-incompatible arrangements (gray open circles) and
  83 EC-compatible structures (blue); Spearman $\rho = 0.982$,
  Pearson $r = 0.986$, MAE 88~meV over a 2.5~eV range.  The
  EC-compatible configurations concentrate at the low-energy end.
  (b)~The cross-class set: all 24 $\ntil$-max candidate
  configurations (filled circles) and the MLIP-predicted Class-2
  minimum (open circle), spanning the 14 Ga placement classes
  (Sec.~\ref{sec:dft}; the DFT axis shows the mixing enthalpy
  $E_\mathrm{mix}$, the MLIP axis the relative energy from the most
  stable configuration of the set);
  $\rho = 0.946$, $r = 0.991$, MAE 39~meV over a 655~meV range.
  Inset: the low-energy region.}
\end{figure*}

\begin{figure*}[tb]
\includegraphics[width=\textwidth]{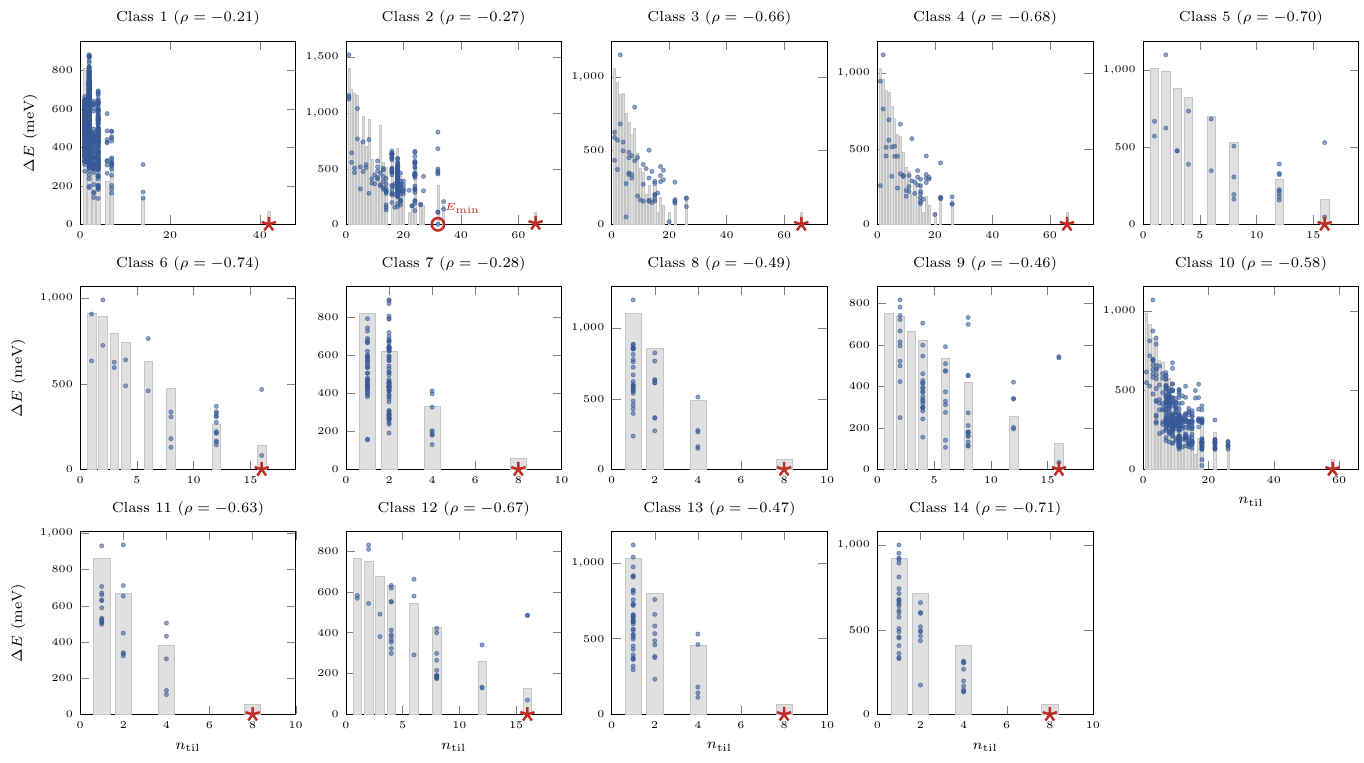}
\caption{\label{fig:ntil_energy}
  MLIP energy versus tiling decomposition multiplicity $\ntil$ for
  the 2,529 relaxed configurations, resolved by Ga placement class
  (one panel per class; energies relative to the class minimum;
  panel titles give the within-class Spearman rank correlation
  $\rho(\ntil, E)$).  The red star in each panel marks the most
  stable configuration among those attaining the class maximum of
  $\ntil$; in 13 of the 14 classes it coincides with the class
  minimum.  The only exception is Class~2, where the class minimum
  (open circle, $\ntil = 32$) lies 8.5~meV below the $\ntil$-max
  configuration.  The light gray bars in the background indicate the
  abundance of each $\ntil$ value among \emph{all} catalog
  configurations of the class; bar heights are proportional to
  $\log(1 + N)$ and normalized panel by panel, and do \emph{not}
  refer to the energy axis.  The density of the plotted points, in
  contrast, reflects the stratified selection of the relaxed sample
  (Sec.~\ref{sec:mlip}), not the catalog abundance.  Both the $\ntil$ range and
the
  $\Delta E$ range differ from panel to panel; the accessible range
  of $\ntil$ is a property of the class
  (Table~\ref{tab:ga_orbits}).}
\end{figure*}

The 416,683 EC-compatible configurations are distributed across the
14 Ga placement classes introduced in Sec.~\ref{sec:enumeration}
(Table~\ref{tab:ga_orbits}; canonical representatives in
Fig.~\ref{fig:ga_classes}).  The 456 tilings themselves also fall
into symmetry classes, which the enumeration uses but which play no
part in the results below (Appendix~\ref{app:symmetry}).
Figure~\ref{fig:ntil_hist} shows the distribution of $\ntil$ across
all configurations.  Class sizes $|O|$ range from $|O| = 12$
(Class~1, whose three Ga adatoms lie on a single lattice line at the
largest attainable spacing, $d^2 = (12,12,12)$) to $|O| = 216$ (the
seven classes whose placements are left unchanged only by the
identity operation of $\Gprime$).  The largest class by pattern
count is Class~10 with 122,317 configurations (29.4\% of the total).

The distribution of $\ntil$ is strongly right-skewed: 256,737
configurations (61.6\%) have $\ntil = 1$, meaning their EC
decomposition is unique, and the mean is 1.75.  The accessible range
of $\ntil$ differs strongly between classes
(Table~\ref{tab:ga_orbits}): the maximum $\ntil = 66$ is attained
only in the three classes whose Ga adatoms are pairwise at the
smallest attainable separation $2a$
[$d^2 = (4,4,4)$, Classes~2--4, one configuration each], whereas in
five classes (7, 8, 11, 13, and~14) $\ntil$ never exceeds~8; $\ntil$ values are
therefore compared only
within a class (Sec.~\ref{sec:ntil_rule}).

\subsection{MLIP--DFT comparison}
\label{sec:mlip_dft}

Figure~\ref{fig:mlip_dft} compares
MLIP and DFT relative energies for the two validation sets of
Sec.~\ref{sec:dft}.

For the fixed-placement set (685 structures at a fixed Class-7 Ga
placement,
including 602 EC-incompatible arrangements), the rank ordering is
preserved to high accuracy across the full 2.5~eV range: the
Spearman correlation is $\rho = 0.982$, the Pearson correlation is
$r = 0.986$, and the MAE of relative energies is 88~meV.  Restricted to the
low-energy region, the ordering remains strong
($\rho = 0.94$ for $\Delta E_\mathrm{DFT} \le 0.6$~eV, $n = 362$;
$\rho = 0.85$ for $\Delta E_\mathrm{DFT} \le 0.3$~eV, $n = 116$).
The MLIP systematically underestimates the destabilization of
high-energy structures: in Fig.~\ref{fig:mlip_dft}(a) the points lie
above the diagonal by a margin that widens with energy, reaching
0.19~eV on average above $\Delta E_\mathrm{DFT} = 1$~eV.  This error
pattern is conservative for the present screening purpose, since
high-energy structures are never stability candidates.

For the cross-class set (25 configurations spanning all 14
classes), the agreement
is $r = 0.991$, $\rho = 0.946$, with an MAE of 39~meV over the
655~meV range [Fig.~\ref{fig:mlip_dft}(b)] and 16~meV in the
low-energy region ($\Delta E_\mathrm{DFT} < 200$~meV, $n = 19$).
The
property that matters most for the analysis that follows is the
preservation of energy ordering within each class, and this is
satisfied exactly.  In all five classes sampled with two or more
configurations (Classes~2, 5, 6, 9, and~12), the MLIP and DFT
energy orderings coincide for every pair of configurations.  The
Class~2 pair included in the cross-class set probes the only
exception to the
$\ntil$-max rule and is discussed in Sec.~\ref{sec:ntil_rule}.

The deviation from the diagonal in Fig.~\ref{fig:mlip_dft} is
dominated by a uniform scale factor.  A linear fit gives
$\Delta E_\mathrm{DFT} \approx 1.17\,\Delta E_\mathrm{MLIP}$ for
the fixed-placement set (slopes 1.16 and 1.17 in its two sampling
runs) and $\approx 1.18$ for the cross-class set, although the two sets employ
different
DFT setups.  The MLIP compresses relative energies by about 15\%,
but it does so uniformly.  Rescaling by this factor reduces the MAE
of the fixed-placement set from 88 to 53~meV (37~meV for
$\Delta E_\mathrm{DFT} \le 0.3$~eV) and that of the cross-class set
from 39 to 27~meV.  Because a positive linear rescaling leaves every energy
ordering unchanged, none of the rank-based results of this paper
depend on this correction; we therefore quote uncalibrated MLIP
energies throughout, and note the scale factor for quantitative use
of the MLIP energy differences.

\subsection{The $\ntil$-max rule}
\label{sec:ntil_rule}

\begin{table}[tb]
\caption{\label{tab:ntilmax}%
Test of the $\ntil$-max rule in all 14 Ga placement classes.
For each class: the size $n$ of the relaxed sample, the identifier
(config ID) of the lowest-energy configuration among the relaxed
configurations of the class, its
multiplicity $\ntil$ over the class maximum $\ntil^\mathrm{max}$
(taken over the full catalog; the relaxed sample contains
configurations at this value in every class), and the within-class
Spearman rank correlation $\rho(\ntil, E)$.  The rule holds when
the two multiplicities coincide, which is the case in 13 of the 14
classes; the only exception is Class~2.  Config IDs (typewriter
type) refer to the openly available catalog
(Sec.~\ref{sec:enumeration}).}
\begin{ruledtabular}
\begin{tabular}{lrrcr}
Class & $n$ & config ID & $\ntil/\ntil^\mathrm{max}$ &
$\rho(\ntil, E)$ \\
\hline
1 & 1,344 & \texttt{6206} & 42/42 & $-0.21$ \\
2 & 227 & \texttt{21900}\rlap{\textsuperscript{a}} & 32/66 & $-0.27$ \\
3 & 54 & \texttt{40652} & 66/66 & $-0.66$ \\
4 & 53 & \texttt{58857} & 66/66 & $-0.68$ \\
5 & 25 & \texttt{70837} & 16/16 & $-0.70$ \\
6 & 29 & \texttt{85176} & 16/16 & $-0.74$ \\
7 & 122 & \texttt{109117} & 8/8 & $-0.28$ \\
8 & 42 & \texttt{130922} & 8/8 & $-0.49$ \\
9 & 65 & \texttt{154062} & 16/16 & $-0.46$ \\
10 & 408 & \texttt{208390} & 58/58 & $-0.58$ \\
11 & 26 & \texttt{313607} & 8/8 & $-0.63$ \\
12 & 40 & \texttt{334746} & 16/16 & $-0.67$ \\
13 & 51 & \texttt{373896} & 8/8 & $-0.47$ \\
14 & 43 & \texttt{413759} & 8/8 & $-0.71$ \\
\end{tabular}
\end{ruledtabular}
\vspace{2pt}
{\raggedright\footnotesize \textsuperscript{a}\,Only exception: the best
$\ntil$-max configuration (\texttt{21882}, $\ntil = 66$) lies 8.5~meV above the
class minimum.\par}
\end{table}

Figure~\ref{fig:ntil_energy} shows the MLIP energy as a function of
$\ntil$ for the 2,529 relaxed configurations, resolved by Ga
placement class.  Energy tends to decrease with increasing $\ntil$
within every class.  Because the accessible $\ntil$ range depends
on the class (Table~\ref{tab:ga_orbits}), the rule is an
\emph{intra-class} statement.  The 14 class minima lie within
120~meV of the global MLIP minimum, compared with an energy spread
of about 1~eV within each class.

The rule tested in this work is stated as follows:

\medskip
\noindent\textit{%
Within each Ga placement class, the lowest MLIP energy among the
relaxed configurations is attained by a configuration that
maximizes the tiling decomposition multiplicity $\ntil$.}
\medskip

\noindent
We test this rule across all 14 Ga placement classes
(Table~\ref{tab:ntilmax}); the relaxed sample contains the
catalog-wide $\ntil$-max configurations of every class, so the
rule is tested against every maximizer, but only against the
relaxed subset of the other configurations.  The rule
holds in 13 of the 14 classes.  The only exception is
Class~2 ($d^2 = (4,4,4)$, $|O| = 36$), where the most stable
configuration has $\ntil = 32$ (rank~4 in the class's $\ntil$
distribution) rather than $\ntil = 66$ (the class maximum).  The
energy gap between the $\ntil$-max configuration and the actual
minimum is 8.5~meV in the MLIP; the independent DFT calculation of
the same two configurations reproduces the ordering with a gap of
17.5~meV (Sec.~\ref{sec:mlip_dft}), confirming that the exception
is a genuine feature of the energy landscape.  For reference, at
typical MOVPE growth temperatures ($T \sim 1300$~K),
$k_B T \approx 110$~meV, several times the DFT gap, so both
configurations may be thermally accessible.  It does not establish
their populations or thermodynamic equivalence: vibrational free
energies, configurational degeneracies, and kinetic accessibility
are not included (Sec.~\ref{sec:outlook}).

The energy margins, the limits of the DFT validation of the class
minima, and checks of the uneven $\ntil$ coverage of the relaxed
sample are given in
Appendix~\ref{app:robust}.

The global MLIP energy minimum across all classes is configuration \cfg{109117}
in Class~7 ($d^2 = (3,3,12)$, $|O| = 108$), with $\ntil = 8$, the
maximum within that class, consistent with the rule.  This is the
configuration whose eight compatible tilings are shown in
Fig.~\ref{fig:interpretations}.  At the DFT level the class minima
move still closer together: the DFT energies of the 14
MLIP-predicted class minima span only 60~meV, and the lowest is the
Class-2 minimum \cfg{21900}, placed 12~meV below \cfg{109117}.

Beyond the extremal statement of the rule, $\ntil$ correlates with
the energy throughout each class.  The within-class Spearman rank
correlation $\rho(\ntil, E)$ is negative in all 14 classes, with a
median of $-0.60$ and a range of $-0.74$ to $-0.21$
(Table~\ref{tab:ntilmax}).  Globally, over all 2,529 relaxed
configurations, $\rho(\ntil, E) = -0.59$; the negative association
persists when the sampled $\ntil$ values are reweighted
(Appendix~\ref{app:robust}).  We also note that the extremal rule and
the monotonic correlation are distinct statements; in Classes~1
and~7 the overall
correlation is weak ($\rho = -0.21$ and $-0.28$), yet the
$\ntil$-max rule holds strictly.

The fixed-placement DFT database (Sec.~\ref{sec:dft}) provides an
additional test at the DFT level, independent of the MLIP, within
the fixed Class-7 Ga placement.  First, EC compatibility itself
separates the energies.  The most stable structure of the entire
685-structure set is EC-compatible; the median relative energy of
the 83 EC-compatible structures is 0.29~eV, against 0.64~eV for the
602 EC-incompatible ones; and 13 of the 20 lowest-energy structures
are EC-compatible, although EC-compatible structures make up only
12\% of the set.  Second, resolved by multiplicity (where
$\ntil = 0$ denotes the EC-incompatible structures of this set; the
catalog of Sec.~\ref{sec:enumeration} contains only
$\ntil \ge 1$), the median DFT
energy decreases monotonically along $\ntil = 0 \to 1 \to 2 \to 4$:
0.64, 0.40, 0.21, and 0.004~eV, respectively.  This monotonic
trend is obtained entirely at the DFT level, without reference to
the MLIP.  Within the EC-compatible subset (81 distinct
configurations with $\ntil \in \{1, 2, 4\}$), the rank correlation
is $\rho(\ntil, E_\mathrm{DFT}) = -0.45$, and the DFT minimum of the
subset is attained by a configuration with the largest $\ntil$
present: the $\ntil$-max rule at the DFT level, within the sampled
range.  (The subset does not contain the class-maximum $\ntil = 8$
configurations, so this is a test within $\ntil \le 4$.)

\subsection{The mechanism: adjacent bare sites}
\label{sec:nadj}

The energy ordering is largely associated with a local feature:
adjacent bare surface Ga sites.

We call a \utri\ face a \emph{bare site} if it hosts no H atom and
is not adjacent to a Ga adatom; by Eq.~(\ref{eq:ec_line}), every
EC-compatible configuration has exactly 9 bare sites.  A bare site is a topmost
Ga atom whose dangling bond prefers to be emptied by donating its
electrons to acceptor states elsewhere on the surface, and two
adjacent bare sites concentrate such electron-donating sources in
one place.  The Ising model of Ref.~\cite{Kawka2024}, a pairwise
interaction model fitted to DFT energies, assigns a repulsive
interaction to these empty-site pairs.
We therefore count the nearest-neighbor pairs of bare sites,
\begin{equation}
  \nadj(G, H) = \bigl|\{\{b, b'\} : b, b' \text{ bare},
    b' \in \mathcal{N}(b)\}\bigr|,
  \label{eq:nadj}
\end{equation}
where $\mathcal{N}(b)$ denotes the nearest-neighbor \utri\ faces
of $b$.

\begin{figure*}[tb]
\includegraphics[width=\textwidth]{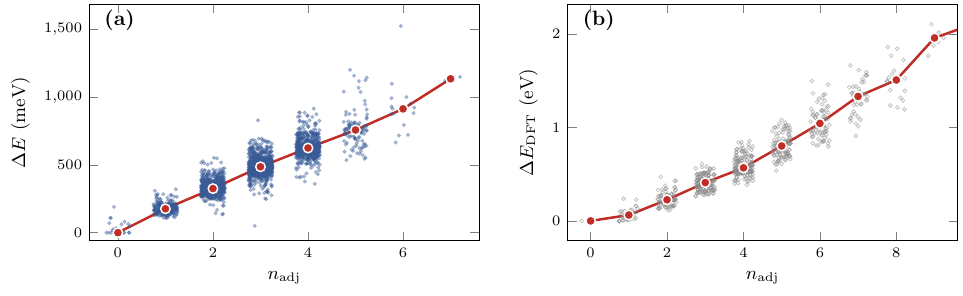}
\caption{\label{fig:vacadj}
  Energy versus the local frustration count $\nadj$
  [Eq.~(\ref{eq:nadj})].
  (a)~MLIP energies of the 2,529 relaxed configurations, relative
  to the respective class minimum; red circles mark the median at
  each $\nadj$.
  (b)~DFT energies of all 685 structures of the fixed-placement
  validation set (Sec.~\ref{sec:dft}), including the 602
  EC-incompatible arrangements, relative to the set minimum;
  red circles mark the median at each $\nadj$.  In both data sets
  the energy increases nearly linearly with $\nadj$, with a cost
  of roughly 0.17~eV (MLIP, within-class median) and 0.22~eV (DFT)
  per adjacent pair.}
\end{figure*}

This pair count tracks the overall energy variation
(Fig.~\ref{fig:vacadj}); cell-averaged descriptors of the same
configurations do not (Sec.~\ref{sec:isotropy}).  The within-class correlation
$\rho(\nadj, E)$ has a median of $+0.92$ (range $+0.86$ to
$+0.98$), and, since $\nadj$ is not a class-relative quantity,
$\rho = +0.92$ over the pooled sample.  The energy rises nearly
linearly with $\nadj$, by 165~meV per adjacent pair (within-class
median; range 131--222~meV).  The DFT data of the fixed-placement
set confirm the relation independently of the MLIP and beyond the
EC-compatible space: $\rho(\nadj, E_\mathrm{DFT}) = +0.96$ over
all 685 structures, 216~meV per pair, and $\rho = +0.84$ to
$+0.93$ within fixed $\ntil$ bins.

To separate the information carried by $\nadj$ and $\ntil$, we
compare configurations with one descriptor held fixed.  At fixed
class and fixed $\ntil$, $\nadj$ orders the energies (median
$\rho = +0.86$ over 51 groups defined by class and $\ntil$, each
with at least six configurations; positive in every group).
At fixed class and fixed $\nadj \geq 1$,
the residual correlation of $\ntil$ essentially vanishes (median
$\rho = -0.05$).

Once adjacent bare-site pairs are absent ($\nadj = 0$), $\ntil$
still distinguishes the energies.  Only 102 of the 416,683
configurations (0.024\%) have $\nadj = 0$; we relaxed all of them
(26 in the original sample and 76 additional relaxations).  Within this
complete set $\ntil$ still orders the energies
[$\rho(\ntil, E) < 0$ in all 14 classes, median $-0.68$], and the
$\ntil$-max configuration is the lowest in 13 of the 14 classes,
by 16--191~meV over the next member of the set; the only exception is again
Class~2 (8.5~meV).
None of the 102 falls below the class minima already identified.
Thus $\nadj$ sets the coarse energy scale, and $\ntil$ orders the
configurations that remain at $\nadj = 0$.  The same division
accounts for the $\ntil$-max candidates that sit far above their
class minima: the six lying 0.47--0.54~eV above are exactly those
with $\nadj = 3$.

The exceptional class is geometrically singular.  Classes~1 and~2
are the only two whose Ga adatoms lie on a single lattice line and
form a chain winding around the torus rather than a triangle, and
Class~2 is the denser of the two, with a spacing of $2a$ against
$2\sqrt{3}\,a$ (Classes~3 and~4 share the signature
$d^2 = (4,4,4)$ with Class~2 but are genuine triangles).

Within Class~2, the exception is resolved by two finer features of the
$\nadj = 0$ set: the number of second-shell bare-site pairs (pair
distance $\sqrt{3}$; $\rho > 0$ in all 14 classes, median
$+0.84$) and the Ga--bare-site proximity
$V_{\mathrm{Gb}} = \sum_{g,b} 1/d_{gb}$ ($\rho < 0$ in all 14
classes, median $-0.82$), again the two interactions of the
Ising model.  The $\ntil$ maximizer, configuration \cfg{21882},
is one of only five members of
the $\nadj = 0$ set whose bare sites avoid even the second shell;
this maximal spreading is what makes $\ntil = 66$ the class
maximum, and the other four such configurations are the minima of
Classes~1, 3, 4, and~10.  The Class-2 minimum, configuration
\cfg{21900}, concedes six
second-shell pairs, places its bare sites closer to the Ga
adatoms, and wins by 8.5~meV (17.5~meV in DFT).  Indeed, the
configuration maximizing $V_{\mathrm{Gb}}$ is the class minimum
in all 14 classes, and this remains true when the untileable
arrangements discussed below are included.  Unlike the
$\ntil$-max rule, this criterion was identified on the same 102
configurations that it describes, and it has not been tested on
other supercells or compositions.

Avoiding adjacent bare sites does not ensure that a compatible
tiling exists.  For example, the $\nadj = 1$ structures of the
DFT set split evenly into EC-compatible and EC-incompatible ones
(17 each), and of the 12 bare-site arrangements with $\nadj = 0$
at the Class-7 placement, one is untileable.

At $\nadj \geq 1$ the
energetic effect of EC compatibility is modest but consistent:
the EC-compatible structures lie 21--72~meV below the
EC-incompatible ones in the median at every populated $\nadj$
(combined Mann--Whitney $z = -4.3$, a rank test for the difference
between two distributions).

The effect is larger among arrangements with $\nadj = 0$.
The canonical representative placements of the 14 classes admit
246 bare-site arrangements with $\nadj = 0$; the 158 tileable ones
reduce, under the operations that leave the placement unchanged, to
the 102
symmetry-distinct configurations of the census, while the
remaining 88 (36\%) are untileable.  We relaxed all 88.  Of these, 83 remained
in their initial adsorption pattern
and lie 68--260~meV (median 146~meV) above their class minima:
near the class minima, the existence of a compatible tiling
corresponds to an energy difference of order 100~meV, comparable
to or exceeding $k_B T$ at growth temperature.

In the remaining
five, a single Ga adatom left its T$_4$ site and settled on an
adjacent bare hollow site---the threefold site between three
surface Ga atoms rather than above an N atom---with all other
displacements below
0.8~\AA; these structures leave the configuration space
considered here, and their MLIP energies (0.1--0.3~eV below the
global minimum of the T$_4$-based space) lie in a region not
validated against DFT.  A DFT examination of this Ga-on-hollow
motif is left for future work.

The two descriptors are complementary.  Enumeration supplies all
$\ntil$ maximizers, while $\nadj$ explains the coarse energy
ordering; within the complete $\nadj = 0$ set, $\ntil$ retains
its association with stability.

\subsection{Cell-averaged descriptors do not capture the ordering}
\label{sec:isotropy}

We next test whether cell-averaged geometry or tiling orientation
captures the same energy ordering.  We consider three descriptors.  The
H--H repulsion proxy $\VHH = \sum_{i<j} 1/d_{ij}$ (summed over the
153 H pairs, with distances between face centers taken in the
minimum-image convention, i.e., along the shortest path allowed by
the periodic boundary conditions)
measures how evenly the H atoms are spread.  The mean Ga--H
separation $\dGaH$ measures the segregation of Ga adatoms from the H
domain.  Finally, the \emph{tiling orientational isotropy} $\Iiso$
aggregates the orientation distributions of all compatible tilings:
with the aggregate orientation fractions
\begin{equation}
  \bar{f}_d
  = \frac{1}{9\,\ntil} \sum_{i=1}^{\ntil} n_d^{(i)},
  \quad d \in \{1, 2, 3\},
\end{equation}
we define
\begin{equation}
  \Iiso = 1 - \frac{3}{2}\sum_{d}
    \left(\bar{f}_d - \frac{1}{3}\right)^{\!2},
  \label{eq:Iiso}
\end{equation}
which equals~1 for a perfectly isotropic aggregate distribution and
decreases with increasing anisotropy.

None of these descriptors explains the stability ranking over the
full relaxed data set.  Over the 2,529 configurations, the Pearson
correlations with the MLIP energy are $r = +0.04$ for $\Iiso$
($+0.05$ for a Shannon-entropy variant of the same quantity),
$r = -0.14$ for $\VHH$, and $r = -0.31$ for $\dGaH$; an ordinary
least-squares model using all three descriptors reaches only
$R^2 = 0.31$.  Within individual classes the correlations are
equally inconsistent (e.g., $r(\Iiso, E)$ ranges from $-0.53$ to
$+0.08$ across the 14 classes).  We note that on small,
stability-enriched subsets of configurations, such as those
selected by ranking heuristics in preliminary stages of this work,
$\Iiso$ appears strongly correlated with the energy; this
correlation is an artifact of the selection and disappears on the
stratified sample.

The within-class ordering is carried by the enumerative
invariant $\ntil$ (Sec.~\ref{sec:ntil_rule}) and by the local count
$\nadj$ (Sec.~\ref{sec:nadj}), but not by cell-averaged descriptors
of the same configurations.  Why $\ntil$ carries information that
a cell average does not is taken up in Sec.~\ref{sec:physics}.

\section{Discussion}
\label{sec:discussion}

\subsection{Why the multiplicity works}
\label{sec:physics}

A compatible tiling decomposes the global EC constraint into nine
local balance equations, one per rhombus block.  The multiplicity
$\ntil$ counts how many such decompositions exist.  Both the
orientations available to Ga blocks and the reassignment of faces
not fixed by the adatoms contribute to this count
(Appendix~\ref{app:symmetry}).

The decreasing median DFT energy along $\ntil = 0 \to 1 \to 2 \to 4$
in the fixed-placement set (Sec.~\ref{sec:ntil_rule}) supports an
association between stability and the ways local EC balance can be
satisfied.  At $\ntil = 0$, no decomposition into locally balanced
blocks exists; at $\ntil = 1$, it is unique.  Additional tilings
provide alternative decompositions of the same atomic configuration,
although they need not differ around every adatom: two tilings can
assign the same blocks to all three Ga adatoms and differ only in
the H blocks.

Unlike cell averages, which can obscure unfavorable local motifs,
$\ntil$ depends on whether local assignments can form a complete
decomposition.  It is a global count of mutually compatible
assignments, not a product of independent local factors.  A motif
with only one local block assignment restricts the possible
decompositions without by itself
reducing $\ntil$ to 1.  An adjacent pair of bare sites is a different
kind of defect: it raises the energy but does not by itself prevent
a decomposition.  Thus some configurations with $\nadj > 0$ still
attain the class maximum of $\ntil$ (Sec.~\ref{sec:nadj}).

The $\nadj$ analysis explains much of the association between
$\ntil$ and energy through the cost of adjacent bare sites, while
$\ntil$ retains information within the $\nadj = 0$ set.
An explanation based on tiling orientational isotropy is not
supported by the data (Sec.~\ref{sec:isotropy}).

\subsection{Structural parallel with the Kekul\'{e} structure count}
\label{sec:kekule}

The relationship between $\ntil$ and stability is structurally
analogous to the correlation between the Kekul\'{e} structure count
$K$---the number of ways double bonds can be assigned so that every
carbon atom carries exactly one, a perfect matching of the carbon
skeleton---and thermodynamic stability in polycyclic aromatic
hydrocarbons.
In molecular chemistry, $K$ has been used since the 1970s as a
quantitative predictor of aromatic stability:
Swinborne-Sheldrake, Herndon, and Gutman~\cite{Swinborne1975}
showed that the Dewar resonance energy of
benzenoid hydrocarbons is proportional to $\ln K$, and later
work, reviewed in Ref.~\cite{Randic2003}, found that this
correlation holds within classes of structural isomers sharing the
same molecular formula.  The present work extends this combinatorial
perspective to a
surface science context, where the counting target changes from
double-bond assignments to EC-compatible tiling decompositions.
Both $K$ and $\ntil$ count the number of ways a global
constraint (electron pairing in molecules; EC balance on a surface)
can be satisfied by local structural units (double bonds; rhombus
blocks), and in both cases the correlation with stability is an
\emph{intra-class} phenomenon.  The key difference is that $K$
reflects quantum-mechanical resonance through the superposition of
valence-bond structures, whereas $\ntil$ is a classical combinatorial
quantity whose link to stability we traced to the avoidance of
adjacent bare sites (Secs.~\ref{sec:nadj} and~\ref{sec:physics}).  The parallel
extends to the
functional form.  Over the 2,529 relaxed configurations, the linear
correlation of the energy is slightly stronger with $\ln \ntil$
($r = -0.60$) than with $\ntil$ itself ($r = -0.54$), a small
difference, but consistent with a logarithmic tendency analogous to
the relation $\mathrm{DRE} \propto \ln K$ of
Ref.~\cite{Swinborne1975}.  The
within-class slope of the energy versus $\ln \ntil$ has a median of
$-195$~meV per $e$-fold of $\ntil$ (range $-75$ to $-300$~meV
across the 14 classes), setting the energy scale associated with
the multiplicity.

\subsection{Comparison with configurational entropy approaches}
\label{sec:entropy}

Mora-Fonz~\textit{et~al.}~\cite{MoraFonz2017} demonstrated that the
polar surfaces of ZnO are stabilized by the large number of distinct
nonstoichiometric configurations satisfying the dipole-cancellation
constraint, via the entropic contribution
$-T S_\mathrm{config} = -k_B T \ln \Omega$ to the free energy.
Duzik~\textit{et~al.}~\cite{DuzikMillunchick2013} found a similar
trend on Bi/GaAs(001).  In both cases the relevant multiplicity
$\Omega$ counts distinct atomic configurations.  Our $\ntil$ is a
different quantity.  It counts decompositions of a \emph{single}
configuration into locally EC-satisfying units.  The $\ntil$-max rule
therefore identifies a mechanism distinct from entropic
stabilization, one operating at the level of structural flexibility
within a given configuration rather than statistical averaging over an
ensemble.  The two mechanisms are nevertheless compatible.  Among the
relaxed sample, 20 configurations drawn from 12 of the 14 classes lie
within $k_B T$ ($T = 1300$~K) of the global minimum, so a thermal
ensemble at growth temperature may include many classes
simultaneously, and the configurational-entropy contribution of
Refs.~\cite{MoraFonz2017,DuzikMillunchick2013} could add to the
single-configuration effect studied here.

\subsection{Practical implications}
\label{sec:practical}

The practical gain is a reduction from 416,683 configurations to
24 $\ntil$-max candidates, between one and four per Ga placement
class (Table~\ref{tab:ga_orbits}).  All 24 have been evaluated with
DFT, together with the MLIP-predicted Class~2 minimum
(Sec.~\ref{sec:dft}).  Equal multiplicity does not imply equal
energy: the three Class~5 maximizers span 0.53~eV in the MLIP and
0.60~eV in DFT.  Energy calculations are therefore still needed to
rank the candidates.

The 25-configuration DFT set contains all MLIP-predicted class
minima and reproduces the Class~2 exception, with a 17.5-meV
inversion (Sec.~\ref{sec:ntil_rule}).  It does not establish the
minima against every excluded competitor (Appendix~\ref{app:robust}),
and the small differences between classes are beyond the resolution
of the present screening.  Whether the candidate set remains small
enough for DFT in larger supercells has not been tested.

The relation to the sampling-based workflow of
Refs.~\cite{Kusaba2022,Kawka2024} is complementary rather than
competing.  Bayesian optimization locates low-energy structures
efficiently, and the data-driven Ising model of
Ref.~\cite{Kawka2024} evaluates an arbitrary number of candidates
at negligible cost, with a reported accuracy of about 0.18~eV
(RMSE), coarser than the 120~meV spread of the class minima found
here.  Both are built on DFT input: sequentially sampled energies
in one case, fitted coupling parameters in the other.  Computing $\ntil$ uses
no energy data, only the adatom geometry
and the 456 tilings.  Enumeration also yields a class-resolved view of the
entire EC-compatible
space (Table~\ref{tab:ga_orbits}, Fig.~\ref{fig:ntil_energy}) and
a descriptor with a physical meaning
(Sec.~\ref{sec:physics}).  The two approaches nevertheless agree
where they overlap.  With the parameters of its reference
implementation~\cite{PyAPX}, the Ising model of
Ref.~\cite{Kawka2024}, evaluated over all 14,896 EC-compatible
configurations at the Class-7 Ga placement, has configuration
\cfg{109117} (the global MLIP minimum identified here) as its unique
minimum,
and $\nadj$ (Sec.~\ref{sec:nadj}) is the discrete
counterpart of its empty-site repulsion, ordering the energies at
fixed class and $\ntil$ more reliably than the fitted model
(median $\rho = +0.86$ versus $+0.77$).  For system sizes beyond
complete enumeration, the two combine naturally, e.g.,
$\ntil$-max candidates as initial data for Bayesian optimization
or as constraints on surrogate training.

\subsection{Limitations and outlook}
\label{sec:outlook}

Several limitations should be noted.  First, the present analysis is
restricted to the $(6 \times 6)$ surface with 3 Ga adatoms and 18 H
atoms ($k_G = 3$).  Our enumeration code supports the other
compositions on the EC line of Eq.~(\ref{eq:ec_line}),
$k_G = 0$--$9$; verifying the rule across compositions is left for
future work.  Selecting the equilibrium composition itself is a
grand-canonical question governed by the chemical potentials of Ga
and H~\cite{Kusaba2017}; the present ranking is at fixed
composition.  Extension to larger supercells, particularly
$(12 \times 12)$, is needed to confirm the generality of the rule
and will require more efficient enumeration, e.g., the ZDD-based
frontier method~\cite{Kawahara2017} or canonical
augmentation~\cite{McKay1998}.
\emph{Applying} the rule, however,
needs only the configurations attaining the largest $\ntil$ in
each placement class; finding these without generating the whole
configuration space, for instance by enumerating the $\nadj = 0$
subspace first (Sec.~\ref{sec:nadj}), is an open combinatorial
problem.  The complete enumeration performed here had a different
purpose, identifying all descriptor maximizers before testing their
energetic preference on selected configurations.

Second, the MLIP--DFT comparison indicates a mean absolute
deviation of relative energies of order 20~meV in the low-energy
region (Sec.~\ref{sec:mlip_dft}), which limits the resolution of
stability ranking for near-degenerate configurations.  More
extensive DFT validation, particularly for the near-degenerate
cases at the top of each class, would better resolve their ordering.
Vibrational free-energy differences could reorder minima separated
by tens of meV at growth temperature.  Configurational degeneracy
(the number of symmetry-equivalent copies) and kinetic accessibility
during growth are also excluded from the present comparison of
single-configuration energies.  Comparisons with $k_B T$
(Secs.~\ref{sec:ntil_rule} and~\ref{sec:entropy}) therefore suggest
possible thermal accessibility, not populations or thermodynamic
equivalence.

Both DFT data sets use GGA functionals: PBE for the fixed-placement
set and PBEJsJrLO for the cross-class set, consistent with earlier
studies of this surface~\cite{Kusaba2017,Kusaba2022,Kawka2024}.
Semilocal functionals underestimate the GaN band gap.  Some functional
error is expected to cancel between configurations of the same
composition on the same slab, but we have not quantified this
cancellation.  Hybrid-functional calculations exist for bare and
hydrogenated GaN surfaces~\cite{Dreyer2014}, but that study does not
test the $(6\times6)$ configurations compared here.  Our energy
ordering remains a GGA-level result; agreement between the MLIP and
DFT does not test functional sensitivity because both rest on GGA
data.  A higher-rung check is left for future work, especially for
the seven classes with margins of 16--48~meV
(Appendix~\ref{app:robust}) and the Class~2 inversion.

Third, the $\ntil$-max rule is an empirical observation on the
$(6 \times 6)$ surface, not a theorem.  A mathematical proof, or a
counterexample in larger systems, would clarify its range of
validity.

Fourth, the predictions have not been confronted with experiment.
The bare-site patterns of the predicted class minima translate into
a $(6 \times 6)$-periodic corrugation that is in principle
accessible to scanning tunneling
microscopy~\cite{Feenstra1998,Bermudez2017}, but atomic-resolution
observation under MOVPE conditions is difficult, and no such
comparison is attempted here.

Finally, the microscopic mechanism linking $\ntil$ to stability is
a reading supported by the monotonic DFT trend, the
$\nadj$ analysis of Sec.~\ref{sec:nadj}, and the local-constraint
picture of Sec.~\ref{sec:physics}, not a demonstrated causal
chain.  No electronic-structure analysis (densities of states or
charge partitioning) was performed; the electron-donation picture
rests on the EC rule and on the energetics.

\section{Conclusions}
\label{sec:conclusions}

Tiling decomposition multiplicity provides a structural descriptor
for screening GaN(0001)-$(6 \times 6)$ configurations at fixed
composition.  Among 2,529 configurations relaxed with a DFT-validated
MLIP, a maximizer of $\ntil$ attains the lowest energy in 13 of the
14 Ga placement classes.  The Class~2 exception is a
high-multiplicity configuration lying 8.5~meV below the maximizer
(17.5~meV in DFT).

The energy ordering is largely associated with avoiding adjacent
bare surface Ga sites: each pair costs about 0.2~eV.  Among
configurations without such pairs, compatible tile decompositions
and their multiplicity provide further information on stability
that the cell-averaged descriptors do not capture.

The complete catalog identifies all 24 $\ntil$ maximizers among
416,683 configurations; these candidates and the Class~2 exception
have been evaluated with DFT.  The energy ranking remains tested
on the relaxed subset, so an unevaluated configuration could still
lie lower.  Within this scope, counting locally balanced
decompositions supplies a small, reproducible candidate set for
first-principles evaluation without requiring an energy calculation
for every configuration.

\begin{acknowledgments}
This work was supported by JSPS KAKENHI Grant No.~23K28151,
by JST BOOST Grant No.~JPMJBY24C3,
and by the Collaborative Research Program of the Research Institute
for Applied Mechanics, Kyushu University.
We gratefully acknowledge Polish high-performance computing
infrastructure and the Interdisciplinary Centre for Mathematical and
Computational Modelling, University of Warsaw (ICM UW), for providing
computer facilities and support within computational Grant
No.~g101-2348.
\end{acknowledgments}

\par\medskip
\noindent
T.K. and A.K. conceived the study.  T.K. developed the tiling
formalism, performed the enumeration and the MLIP screening, and
wrote the first draft of the manuscript.  K.K. and P.K. performed
the DFT calculations.  All authors discussed the results and
reviewed and revised the manuscript.
\par\medskip

\section*{Data availability}
The configuration catalog (416,683 configurations with their
identifiers, adatom coordinates, and $\ntil$ values), the
relaxed-energy data, the DFT--MLIP comparison data, and the
enumeration and analysis scripts that support the findings of this
article are openly available in Zenodo~\cite{GaNTilingData2026}.
The configuration identifiers used in this paper refer to the
\texttt{config\_id} column of the released catalog.

\appendix

\section{Symmetry classification and enumeration details}
\label{app:symmetry}

This appendix collects the group-theoretic and algorithmic material
behind the lattice symmetry of Sec.~\ref{sec:lattice} and the
enumeration of Sec.~\ref{sec:enumeration}, and closes with the
combinatorial anatomy of $\ntil$ referred to in
Sec.~\ref{sec:physics}.

\subsection{The symmetry group}

\begin{figure}[tb]
\includegraphics[width=\columnwidth]{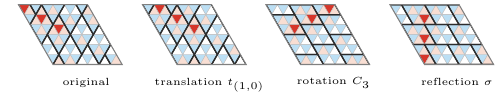}
\caption{\label{fig:symmetry}
  Symmetry operations of the sublattice-preserving group
  $\Gprime = T \rtimes \Dthree$.  The leftmost panel reproduces
  the schematic view of Fig.~\ref{fig:lattice}(c); the other panels
  show its images under a translation, a $120^\circ$ rotation, and
  a reflection, the three types of generators of $\Gprime$.  Each
  operation acts simultaneously on the block boundaries and on the
  adatom positions.  Configurations related by any of the 216
  elements of $\Gprime$ are physically identical and are counted
  once in the enumeration (Sec.~\ref{sec:enumeration}).}
\end{figure}

The full symmetry group of the lattice is
$\Gfull = T \rtimes \Dsix$ with $|\Gfull| = 432$, the semidirect
product ($\rtimes$) of the group
$T \cong \mathbb{Z}_6 \times \mathbb{Z}_6$ of the 36 torus
translations and the 12-element dihedral point group $\Dsix$ of the
triangular lattice.
Because Ga adatoms and H atoms occupy distinct sublattices
(\dtri\ and \utri, respectively), the operations in $\Dsix$ that
exchange the two sublattices are physically irrelevant.  The
sublattice-preserving subgroup is
$\Gprime = T \rtimes \Dthree$ with $|\Gprime| = 216$, where
$\Dthree$ retains the rotations by multiples of $120^\circ$ and the
three reflections that map each sublattice to itself; this is the
symmetry group used throughout to classify configurations.
Figure~\ref{fig:symmetry} illustrates the three types of generators
of $\Gprime$ acting on the tiled configuration of
Fig.~\ref{fig:lattice}(c).

\subsection{Tiling classes}

\begin{figure*}[tb]
\centering
\includegraphics{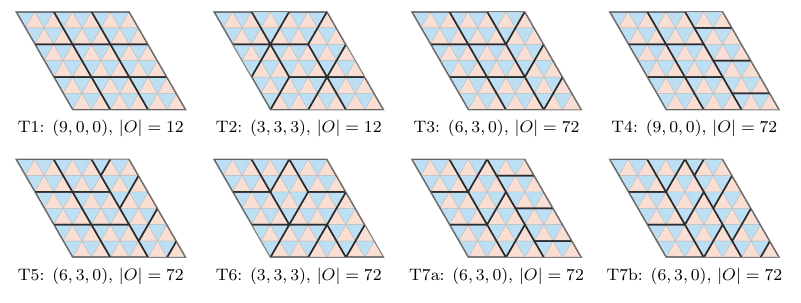}
\caption{\label{fig:tiling_classes}
  Representatives of the 8 tiling classes of the
  adatom-free $(6 \times 6)$ torus under the sublattice-preserving group
  $\Gprime$ (Table~\ref{tab:tiling_classes}), labeled by the sorted
  orientation multiset and the orbit size $|O|$.  No adatoms are
  involved at this stage.  The pairs T1/T4 and T2/T6 share orientation
  multisets but differ in the relative arrangement of equally
  oriented blocks; T7a and T7b are exchanged by the $60^\circ$
  rotation and are therefore equivalent under the full group $\Gfull$ but
  not under $\Gprime$.  The coloring enumeration of
  Sec.~\ref{sec:enumeration} operates on these 8 representatives
  only.}
\end{figure*}

The 456 rhombus tilings on the $(6 \times 6)$ torus are classified
into 8 equivalence classes under $\Gprime$
(Table~\ref{tab:tiling_classes}; representatives in
Fig.~\ref{fig:tiling_classes}).  Under the full
lattice symmetry $\Gfull$ they fall into 7 equivalence classes; under
$\Gprime$ one class (T7) splits into two orbits, because its
stabilizer in $\Gfull$ (the subgroup of symmetry operations that map
the tiling to itself) contains no sublattice-exchanging element.
We label the two orbits T7a and T7b to record their origin.  They are
exchanged by the sublattice-exchanging operations of $\Gfull$
(e.g., the $60^\circ$ rotation) and are distinct
classes only under $\Gprime$.  This yields 8 physical tiling classes
(Fig.~\ref{fig:tiling_classes} and Table~\ref{tab:tiling_classes}).

The classes differ in their
orientation distributions $(n_1, n_2, n_3)$, which record how many
of the 9 rhombus blocks take each of the three orientations of
Fig.~\ref{fig:pieces}(a); symmetry operations permute the
orientations, so the
multiset of the three counts is a class invariant.  Two classes (T2
and T6, 84 tilings in total) consist of tilings with the maximally
isotropic distribution $(3,3,3)$: in T2 the equally oriented blocks
are mutually non-adjacent, whereas in T6 they are adjacent.  The
remaining classes have biased distributions, of $(6,3,0)$ type or
the fully aligned $(9,0,0)$.

\begin{table}[tb]
\caption{\label{tab:tiling_classes}%
The 456 rhombus tilings of the $(6 \times 6)$ torus classified under
the sublattice-preserving group $\Gprime$.  For each class, the
multiset of the orientation counts $\{n_1, n_2, n_3\}$ (sorted in
decreasing order; orientations as in Fig.~\ref{fig:pieces}(a)), the
orbit size $|O|$, and
the order of the stabilizer subgroup in $\Gprime$,
$|\mathrm{Stab}| = |\Gprime|/|O|$, are listed.  Classes T7a and T7b
are exchanged by the $60^\circ$ rotation and are equivalent under
the full group $\Gfull$ but not under $\Gprime$.}
\begin{ruledtabular}
\begin{tabular}{lcrr}
Class & orientations & $|O|$ & $|\mathrm{Stab}|$ \\
\hline
T1 & $(9,0,0)$ & 12 & 18 \\
T2 & $(3,3,3)$ & 12 & 18 \\
T3 & $(6,3,0)$ & 72 & 3 \\
T4 & $(9,0,0)$ & 72 & 3 \\
T5 & $(6,3,0)$ & 72 & 3 \\
T6 & $(3,3,3)$ & 72 & 3 \\
T7a & $(6,3,0)$ & 72 & 3 \\
T7b & $(6,3,0)$ & 72 & 3 \\
\end{tabular}
\end{ruledtabular}
\end{table}

\subsection{Canonical forms}

Each face carries an index
$\mathrm{face\_id} = 2(6q + r) + t$, with $(q, r)$ axial coordinates
on the torus and $t \in \{0, 1\}$ the sublattice index; this is the
index used in the openly available catalog.

Two canonicalization steps make the reduction of
Sec.~\ref{sec:enumeration} exact.  In Stage~2, restricting the
coloring enumeration to the 8 representative tilings loses nothing,
because every tiling is the $\Gprime$-image of a representative and
colorings transform covariantly under the group action; each
coloring is converted to a position pattern $(G,H)$ and reduced
under the stabilizer subgroup of its representative tiling.
Identical position patterns within and across representatives are
merged, giving 670,415 patterns.  In Stage~3, each position pattern
is reduced under the full group $\Gprime$ by taking
the lexicographically smallest image of $(G, H)$ over all 216
operations; that image serves as the canonical representative of the
orbit.

\subsection{Consistency check}

The counts quoted in Sec.~\ref{sec:enumeration} are exact
combinatorial statements involving no energy model.  As a check, the
orbit counts---47 Ga placement classes, and 8 and 7 tiling classes
under $\Gprime$ and $\Gfull$ respectively---were reproduced
independently with Burnside's lemma, the orbit-counting identity
that expresses the number of orbits as the average number of objects
left fixed by a group element.

\subsection{Combinatorial anatomy of $\ntil$}

Two combinatorial factors contribute to $\ntil$.  The first is the
Ga orientation freedom $\Pfree$: each Ga adatom can participate in
a rhombus block of up to 3 orientations, but H atoms
on the \utri\ faces of a candidate block may exclude some
orientations.  The product
\begin{equation}
  \Pfree = \prod_{g \in G}
  \bigl|\{d : \text{rhombus}(g, d) \text{ is H-free}\}\bigr|
  \label{eq:pfree}
\end{equation}
takes values from 1 to $3^3 = 27$ and correlates with $\ntil$ at
$r = 0.46$ across the relaxed set.
Direction freedom alone, however, is insufficient; 56 of the 2,529
relaxed configurations have the maximal $\Pfree = 27$ and yet
$\ntil = 1$.  The second factor is the reassignment freedom of the
faces not fixed by the adatoms, which we quantify as $\nflex$, the
number of H-free \utri\ faces whose block assignment differs among
the compatible tilings; it correlates with $\ntil$ at $r = 0.76$
(by construction $\nflex = 0$ whenever $\ntil = 1$, so part of this
correlation is definitional).  Neither factor alone determines
$\ntil$; the multiplicity is a global property of the
configuration, which is why its evaluation requires enumeration
rather than a local formula.

\section{Robustness checks of the $\ntil$-max rule}
\label{app:robust}

\emph{Margins against the MLIP uncertainty.}---How robust is the
identification of the class minima against the MLIP uncertainty?
Over all relaxed configurations, including the complete
$\nadj = 0$ set of Sec.~\ref{sec:nadj}, the $\ntil$-max
configuration of each of the 13 rule-conforming classes lies
16--135~meV (median 48~meV) below the most stable configuration
with any smaller $\ntil$.  In six classes the margin (57--135~meV)
exceeds the calibrated MLIP accuracy of Sec.~\ref{sec:mlip_dft};
in the remaining seven (Classes~3, 5, 8, 10, 11, 13, and~14,
margins 16--48~meV) it is comparable to that accuracy, so
the identity of the class minimum remains an MLIP prediction.  The
DFT candidate set does not contain the nearest lower-$\ntil$
competitors in these seven classes, so the DFT evaluation of the
maximizers alone does not settle those comparisons
(Sec.~\ref{sec:practical}).  With one
exception (Class~1, where the nearest competitor has $\nadj = 2$),
the closest competitors are low-$\ntil$ members of the
$\nadj = 0$ set, so an inversion due to MLIP error would
displace the minimum within the unfrustrated subset rather than
overturn the qualitative picture.

\emph{Sampling density.}---Because the relaxed
sample was selected in part by a score involving $\ntil$
(Sec.~\ref{sec:mlip}), we checked the sensitivity of the within-class
correlations of
Sec.~\ref{sec:ntil_rule} to the weighting of the sampled
$\ntil$ values.  Collapsing each
class to one point per $\ntil$ value (the median energy at each
value) gives a median within-class correlation of $\rho = -0.89$
(range $-1.00$ to $-0.72$), and resampling one random configuration
per $\ntil$ value yields consistently negative correlations.  The negative
association therefore persists under these
reweightings of the selected sample; this check does not assess
configurations outside that sample.


\begin{thebibliography}{36}%
\makeatletter
\providecommand \@ifxundefined [1]{%
 \@ifx{#1\undefined}
}%
\providecommand \@ifnum [1]{%
 \ifnum #1\expandafter \@firstoftwo
 \else \expandafter \@secondoftwo
 \fi
}%
\providecommand \@ifx [1]{%
 \ifx #1\expandafter \@firstoftwo
 \else \expandafter \@secondoftwo
 \fi
}%
\providecommand \natexlab [1]{#1}%
\providecommand \enquote  [1]{``#1''}%
\providecommand \bibnamefont  [1]{#1}%
\providecommand \bibfnamefont [1]{#1}%
\providecommand \citenamefont [1]{#1}%
\providecommand \href@noop [0]{\@secondoftwo}%
\providecommand \href [0]{\begingroup \@sanitize@url \@href}%
\providecommand \@href[1]{\@@startlink{#1}\@@href}%
\providecommand \@@href[1]{\endgroup#1\@@endlink}%
\providecommand \@sanitize@url [0]{\catcode `\\12\catcode `\$12\catcode
  `\&12\catcode `\#12\catcode `\^12\catcode `\_12\catcode `\%12\relax}%
\providecommand \@@startlink[1]{}%
\providecommand \@@endlink[0]{}%
\providecommand \url  [0]{\begingroup\@sanitize@url \@url }%
\providecommand \@url [1]{\endgroup\@href {#1}{\urlprefix }}%
\providecommand \urlprefix  [0]{URL }%
\providecommand \Eprint [0]{\href }%
\providecommand \doibase [0]{https://doi.org/}%
\providecommand \selectlanguage [0]{\@gobble}%
\providecommand \bibinfo  [0]{\@secondoftwo}%
\providecommand \bibfield  [0]{\@secondoftwo}%
\providecommand \translation [1]{[#1]}%
\providecommand \BibitemOpen [0]{}%
\providecommand \bibitemStop [0]{}%
\providecommand \bibitemNoStop [0]{.\EOS\space}%
\providecommand \EOS [0]{\spacefactor3000\relax}%
\providecommand \BibitemShut  [1]{\csname bibitem#1\endcsname}%
\let\auto@bib@innerbib\@empty
\bibitem [{\citenamefont {Zywietz}\ \emph {et~al.}(1998)\citenamefont
  {Zywietz}, \citenamefont {Neugebauer},\ and\ \citenamefont
  {Scheffler}}]{Zywietz1998}%
  \BibitemOpen
  \bibfield  {author} {\bibinfo {author} {\bibfnamefont {T.}~\bibnamefont
  {Zywietz}}, \bibinfo {author} {\bibfnamefont {J.}~\bibnamefont
  {Neugebauer}},\ and\ \bibinfo {author} {\bibfnamefont {M.}~\bibnamefont
  {Scheffler}},\ }\bibfield  {title} {\bibinfo {title} {Adatom diffusion at
  {GaN} (0001) and (000$\bar{1}$) surfaces},\ }\href
  {https://doi.org/10.1063/1.121909} {\bibfield  {journal} {\bibinfo  {journal}
  {Appl. Phys. Lett.}\ }\textbf {\bibinfo {volume} {73}},\ \bibinfo {pages}
  {487} (\bibinfo {year} {1998})}\BibitemShut {NoStop}%
\bibitem [{\citenamefont {Van~de Walle}\ and\ \citenamefont
  {Neugebauer}(2002)}]{VanDeWalleJVST2002}%
  \BibitemOpen
  \bibfield  {author} {\bibinfo {author} {\bibfnamefont {C.~G.}\ \bibnamefont
  {Van~de Walle}}\ and\ \bibinfo {author} {\bibfnamefont {J.}~\bibnamefont
  {Neugebauer}},\ }\bibfield  {title} {\bibinfo {title} {Role of hydrogen in
  surface reconstructions and growth of {GaN}},\ }\href
  {https://doi.org/10.1116/1.1491545} {\bibfield  {journal} {\bibinfo
  {journal} {J. Vac. Sci. Technol. B}\ }\textbf {\bibinfo {volume} {20}},\
  \bibinfo {pages} {1640} (\bibinfo {year} {2002})}\BibitemShut {NoStop}%
\bibitem [{\citenamefont {Neugebauer}\ \emph {et~al.}(2003)\citenamefont
  {Neugebauer}, \citenamefont {Zywietz}, \citenamefont {Scheffler},
  \citenamefont {Northrup}, \citenamefont {Chen},\ and\ \citenamefont
  {Feenstra}}]{Neugebauer2003}%
  \BibitemOpen
  \bibfield  {author} {\bibinfo {author} {\bibfnamefont {J.}~\bibnamefont
  {Neugebauer}}, \bibinfo {author} {\bibfnamefont {T.~K.}\ \bibnamefont
  {Zywietz}}, \bibinfo {author} {\bibfnamefont {M.}~\bibnamefont {Scheffler}},
  \bibinfo {author} {\bibfnamefont {J.~E.}\ \bibnamefont {Northrup}}, \bibinfo
  {author} {\bibfnamefont {H.}~\bibnamefont {Chen}},\ and\ \bibinfo {author}
  {\bibfnamefont {R.~M.}\ \bibnamefont {Feenstra}},\ }\bibfield  {title}
  {\bibinfo {title} {Adatom kinetics on and below the surface: {The} existence
  of a new diffusion channel},\ }\href
  {https://doi.org/10.1103/PhysRevLett.90.056101} {\bibfield  {journal}
  {\bibinfo  {journal} {Phys. Rev. Lett.}\ }\textbf {\bibinfo {volume} {90}},\
  \bibinfo {pages} {056101} (\bibinfo {year} {2003})}\BibitemShut {NoStop}%
\bibitem [{\citenamefont {Akiyama}\ and\ \citenamefont
  {Kawamura}(2024)}]{Akiyama2024CGD}%
  \BibitemOpen
  \bibfield  {author} {\bibinfo {author} {\bibfnamefont {T.}~\bibnamefont
  {Akiyama}}\ and\ \bibinfo {author} {\bibfnamefont {T.}~\bibnamefont
  {Kawamura}},\ }\bibfield  {title} {\bibinfo {title} {Structural stability of
  vicinal {AlN}(0001) and {GaN}(0001) surfaces with steps and kinks under
  metal--organic vapor-phase epitaxy condition: {A} first-principles study},\
  }\href {https://doi.org/10.1021/acs.cgd.4c00121} {\bibfield  {journal}
  {\bibinfo  {journal} {Cryst. Growth Des.}\ }\textbf {\bibinfo {volume}
  {24}},\ \bibinfo {pages} {5906} (\bibinfo {year} {2024})}\BibitemShut
  {NoStop}%
\bibitem [{\citenamefont {Kangawa}\ \emph {et~al.}(2025)\citenamefont
  {Kangawa}, \citenamefont {Kusaba}, \citenamefont {Kawamura}, \citenamefont
  {Kempisty}, \citenamefont {Ishisone},\ and\ \citenamefont
  {Boero}}]{Kangawa2025}%
  \BibitemOpen
  \bibfield  {author} {\bibinfo {author} {\bibfnamefont {Y.}~\bibnamefont
  {Kangawa}}, \bibinfo {author} {\bibfnamefont {A.}~\bibnamefont {Kusaba}},
  \bibinfo {author} {\bibfnamefont {T.}~\bibnamefont {Kawamura}}, \bibinfo
  {author} {\bibfnamefont {P.}~\bibnamefont {Kempisty}}, \bibinfo {author}
  {\bibfnamefont {K.}~\bibnamefont {Ishisone}},\ and\ \bibinfo {author}
  {\bibfnamefont {M.}~\bibnamefont {Boero}},\ }\bibfield  {title} {\bibinfo
  {title} {Influence of intrinsic point defects incorporated from growth
  surface on atomic interdiffusion and unintentional compositional gradient in
  {AlGaN}/{AlN} heterointerfaces},\ }\href
  {https://doi.org/10.1021/acs.cgd.4c01542} {\bibfield  {journal} {\bibinfo
  {journal} {Cryst. Growth Des.}\ }\textbf {\bibinfo {volume} {25}},\ \bibinfo
  {pages} {740} (\bibinfo {year} {2025})}\BibitemShut {NoStop}%
\bibitem [{\citenamefont {Kusaba}\ \emph {et~al.}(2022)\citenamefont {Kusaba},
  \citenamefont {Kangawa}, \citenamefont {Kuboyama},\ and\ \citenamefont
  {Oshiyama}}]{Kusaba2022}%
  \BibitemOpen
  \bibfield  {author} {\bibinfo {author} {\bibfnamefont {A.}~\bibnamefont
  {Kusaba}}, \bibinfo {author} {\bibfnamefont {Y.}~\bibnamefont {Kangawa}},
  \bibinfo {author} {\bibfnamefont {T.}~\bibnamefont {Kuboyama}},\ and\
  \bibinfo {author} {\bibfnamefont {A.}~\bibnamefont {Oshiyama}},\ }\bibfield
  {title} {\bibinfo {title} {Exploration of a large-scale reconstructed
  structure on {GaN}(0001) surface by {Bayesian} optimization},\ }\href
  {https://doi.org/10.1063/5.0078660} {\bibfield  {journal} {\bibinfo
  {journal} {Appl. Phys. Lett.}\ }\textbf {\bibinfo {volume} {120}},\ \bibinfo
  {pages} {021602} (\bibinfo {year} {2022})}\BibitemShut {NoStop}%
\bibitem [{\citenamefont {Kawka}\ \emph {et~al.}(2024)\citenamefont {Kawka},
  \citenamefont {Kempisty}, \citenamefont {Sakowski}, \citenamefont
  {Krukowski}, \citenamefont {Bo\'{c}kowski}, \citenamefont {Bowler},\ and\
  \citenamefont {Kusaba}}]{Kawka2024}%
  \BibitemOpen
  \bibfield  {author} {\bibinfo {author} {\bibfnamefont {K.}~\bibnamefont
  {Kawka}}, \bibinfo {author} {\bibfnamefont {P.}~\bibnamefont {Kempisty}},
  \bibinfo {author} {\bibfnamefont {K.}~\bibnamefont {Sakowski}}, \bibinfo
  {author} {\bibfnamefont {S.}~\bibnamefont {Krukowski}}, \bibinfo {author}
  {\bibfnamefont {M.}~\bibnamefont {Bo\'{c}kowski}}, \bibinfo {author}
  {\bibfnamefont {D.}~\bibnamefont {Bowler}},\ and\ \bibinfo {author}
  {\bibfnamefont {A.}~\bibnamefont {Kusaba}},\ }\bibfield  {title} {\bibinfo
  {title} {Augmentation of the electron counting rule with {Ising} model},\
  }\href {https://doi.org/10.1063/5.0203033} {\bibfield  {journal} {\bibinfo
  {journal} {J. Appl. Phys.}\ }\textbf {\bibinfo {volume} {135}},\ \bibinfo
  {pages} {225302} (\bibinfo {year} {2024})}\BibitemShut {NoStop}%
\bibitem [{\citenamefont {Pashley}(1989)}]{Pashley1989}%
  \BibitemOpen
  \bibfield  {author} {\bibinfo {author} {\bibfnamefont {M.~D.}\ \bibnamefont
  {Pashley}},\ }\bibfield  {title} {\bibinfo {title} {Electron counting model
  and its application to island structures on molecular-beam epitaxy grown
  {GaAs}(001) and {ZnSe}(001)},\ }\href
  {https://doi.org/10.1103/PhysRevB.40.10481} {\bibfield  {journal} {\bibinfo
  {journal} {Phys. Rev. B}\ }\textbf {\bibinfo {volume} {40}},\ \bibinfo
  {pages} {10481} (\bibinfo {year} {1989})}\BibitemShut {NoStop}%
\bibitem [{\citenamefont {Zhang}\ \emph {et~al.}(2006)\citenamefont {Zhang},
  \citenamefont {Wang}, \citenamefont {Xue}, \citenamefont {Zhang},\ and\
  \citenamefont {Zhang}}]{Zhang2006GEC}%
  \BibitemOpen
  \bibfield  {author} {\bibinfo {author} {\bibfnamefont {L.}~\bibnamefont
  {Zhang}}, \bibinfo {author} {\bibfnamefont {E.~G.}\ \bibnamefont {Wang}},
  \bibinfo {author} {\bibfnamefont {Q.~K.}\ \bibnamefont {Xue}}, \bibinfo
  {author} {\bibfnamefont {S.~B.}\ \bibnamefont {Zhang}},\ and\ \bibinfo
  {author} {\bibfnamefont {Z.}~\bibnamefont {Zhang}},\ }\bibfield  {title}
  {\bibinfo {title} {Generalized electron counting in determination of
  metal-induced reconstruction of compound semiconductor surfaces},\ }\href
  {https://doi.org/10.1103/PhysRevLett.97.126103} {\bibfield  {journal}
  {\bibinfo  {journal} {Phys. Rev. Lett.}\ }\textbf {\bibinfo {volume} {97}},\
  \bibinfo {pages} {126103} (\bibinfo {year} {2006})}\BibitemShut {NoStop}%
\bibitem [{\citenamefont {Kangawa}\ \emph {et~al.}(2013)\citenamefont
  {Kangawa}, \citenamefont {Akiyama}, \citenamefont {Ito}, \citenamefont
  {Shiraishi},\ and\ \citenamefont {Nakayama}}]{Kangawa2013}%
  \BibitemOpen
  \bibfield  {author} {\bibinfo {author} {\bibfnamefont {Y.}~\bibnamefont
  {Kangawa}}, \bibinfo {author} {\bibfnamefont {T.}~\bibnamefont {Akiyama}},
  \bibinfo {author} {\bibfnamefont {T.}~\bibnamefont {Ito}}, \bibinfo {author}
  {\bibfnamefont {K.}~\bibnamefont {Shiraishi}},\ and\ \bibinfo {author}
  {\bibfnamefont {T.}~\bibnamefont {Nakayama}},\ }\bibfield  {title} {\bibinfo
  {title} {Surface stability and growth kinetics of compound semiconductors: An
  ab initio-based approach},\ }\href {https://doi.org/10.3390/ma6083309}
  {\bibfield  {journal} {\bibinfo  {journal} {Materials}\ }\textbf {\bibinfo
  {volume} {6}},\ \bibinfo {pages} {3309} (\bibinfo {year} {2013})}\BibitemShut
  {NoStop}%
\bibitem [{\citenamefont {Kusaba}\ \emph {et~al.}(2017)\citenamefont {Kusaba},
  \citenamefont {Kangawa}, \citenamefont {Kempisty}, \citenamefont {Valencia},
  \citenamefont {Shiraishi}, \citenamefont {Kumagai}, \citenamefont
  {Kakimoto},\ and\ \citenamefont {Koukitu}}]{Kusaba2017}%
  \BibitemOpen
  \bibfield  {author} {\bibinfo {author} {\bibfnamefont {A.}~\bibnamefont
  {Kusaba}}, \bibinfo {author} {\bibfnamefont {Y.}~\bibnamefont {Kangawa}},
  \bibinfo {author} {\bibfnamefont {P.}~\bibnamefont {Kempisty}}, \bibinfo
  {author} {\bibfnamefont {H.}~\bibnamefont {Valencia}}, \bibinfo {author}
  {\bibfnamefont {K.}~\bibnamefont {Shiraishi}}, \bibinfo {author}
  {\bibfnamefont {Y.}~\bibnamefont {Kumagai}}, \bibinfo {author} {\bibfnamefont
  {K.}~\bibnamefont {Kakimoto}},\ and\ \bibinfo {author} {\bibfnamefont
  {A.}~\bibnamefont {Koukitu}},\ }\bibfield  {title} {\bibinfo {title}
  {Thermodynamic analysis of (0001) and (000$\bar{1}$) {GaN} metalorganic vapor
  phase epitaxy},\ }\href {https://doi.org/10.7567/JJAP.56.070304} {\bibfield
  {journal} {\bibinfo  {journal} {Jpn. J. Appl. Phys.}\ }\textbf {\bibinfo
  {volume} {56}},\ \bibinfo {pages} {070304} (\bibinfo {year}
  {2017})}\BibitemShut {NoStop}%
\bibitem [{\citenamefont {Hart}\ and\ \citenamefont
  {Forcade}(2008)}]{HartForcade2008}%
  \BibitemOpen
  \bibfield  {author} {\bibinfo {author} {\bibfnamefont {G.~L.~W.}\
  \bibnamefont {Hart}}\ and\ \bibinfo {author} {\bibfnamefont {R.~W.}\
  \bibnamefont {Forcade}},\ }\bibfield  {title} {\bibinfo {title} {Algorithm
  for generating derivative structures},\ }\href
  {https://doi.org/10.1103/PhysRevB.77.224115} {\bibfield  {journal} {\bibinfo
  {journal} {Phys. Rev. B}\ }\textbf {\bibinfo {volume} {77}},\ \bibinfo
  {pages} {224115} (\bibinfo {year} {2008})}\BibitemShut {NoStop}%
\bibitem [{\citenamefont {Herndon}(1973)}]{Herndon1973}%
  \BibitemOpen
  \bibfield  {author} {\bibinfo {author} {\bibfnamefont {W.~C.}\ \bibnamefont
  {Herndon}},\ }\bibfield  {title} {\bibinfo {title} {Resonance energies of
  aromatic hydrocarbons. {Quantitative} test of resonance theory},\ }\href
  {https://doi.org/10.1021/ja00788a073} {\bibfield  {journal} {\bibinfo
  {journal} {J. Am. Chem. Soc.}\ }\textbf {\bibinfo {volume} {95}},\ \bibinfo
  {pages} {2404} (\bibinfo {year} {1973})}\BibitemShut {NoStop}%
\bibitem [{\citenamefont {Swinborne-Sheldrake}\ \emph
  {et~al.}(1975)\citenamefont {Swinborne-Sheldrake}, \citenamefont {Herndon},\
  and\ \citenamefont {Gutman}}]{Swinborne1975}%
  \BibitemOpen
  \bibfield  {author} {\bibinfo {author} {\bibfnamefont {R.}~\bibnamefont
  {Swinborne-Sheldrake}}, \bibinfo {author} {\bibfnamefont {W.~C.}\
  \bibnamefont {Herndon}},\ and\ \bibinfo {author} {\bibfnamefont
  {I.}~\bibnamefont {Gutman}},\ }\bibfield  {title} {\bibinfo {title}
  {Kekul\'{e} structures and resonance energies of benzenoid hydrocarbons},\
  }\href {https://doi.org/10.1016/S0040-4039(00)71975-7} {\bibfield  {journal}
  {\bibinfo  {journal} {Tetrahedron Lett.}\ }\textbf {\bibinfo {volume} {16}},\
  \bibinfo {pages} {755} (\bibinfo {year} {1975})}\BibitemShut {NoStop}%
\bibitem [{\citenamefont {Randi\'{c}}(2003)}]{Randic2003}%
  \BibitemOpen
  \bibfield  {author} {\bibinfo {author} {\bibfnamefont {M.}~\bibnamefont
  {Randi\'{c}}},\ }\bibfield  {title} {\bibinfo {title} {Aromaticity of
  polycyclic conjugated hydrocarbons},\ }\href
  {https://doi.org/10.1021/cr9903656} {\bibfield  {journal} {\bibinfo
  {journal} {Chem. Rev.}\ }\textbf {\bibinfo {volume} {103}},\ \bibinfo {pages}
  {3449} (\bibinfo {year} {2003})}\BibitemShut {NoStop}%
\bibitem [{\citenamefont {Mora-Fonz}\ \emph {et~al.}(2017)\citenamefont
  {Mora-Fonz}, \citenamefont {Lazauskas}, \citenamefont {Farrow}, \citenamefont
  {Catlow}, \citenamefont {Woodley},\ and\ \citenamefont
  {Sokol}}]{MoraFonz2017}%
  \BibitemOpen
  \bibfield  {author} {\bibinfo {author} {\bibfnamefont {D.}~\bibnamefont
  {Mora-Fonz}}, \bibinfo {author} {\bibfnamefont {T.}~\bibnamefont
  {Lazauskas}}, \bibinfo {author} {\bibfnamefont {M.~R.}\ \bibnamefont
  {Farrow}}, \bibinfo {author} {\bibfnamefont {C.~R.~A.}\ \bibnamefont
  {Catlow}}, \bibinfo {author} {\bibfnamefont {S.~M.}\ \bibnamefont
  {Woodley}},\ and\ \bibinfo {author} {\bibfnamefont {A.~A.}\ \bibnamefont
  {Sokol}},\ }\bibfield  {title} {\bibinfo {title} {Why are polar surfaces of
  {ZnO} stable?},\ }\href {https://doi.org/10.1021/acs.chemmater.7b01487}
  {\bibfield  {journal} {\bibinfo  {journal} {Chem. Mater.}\ }\textbf {\bibinfo
  {volume} {29}},\ \bibinfo {pages} {5306} (\bibinfo {year}
  {2017})}\BibitemShut {NoStop}%
\bibitem [{\citenamefont {Rapcewicz}\ \emph {et~al.}(1997)\citenamefont
  {Rapcewicz}, \citenamefont {Buongiorno~Nardelli},\ and\ \citenamefont
  {Bernholc}}]{Rapcewicz1997}%
  \BibitemOpen
  \bibfield  {author} {\bibinfo {author} {\bibfnamefont {K.}~\bibnamefont
  {Rapcewicz}}, \bibinfo {author} {\bibfnamefont {M.}~\bibnamefont
  {Buongiorno~Nardelli}},\ and\ \bibinfo {author} {\bibfnamefont
  {J.}~\bibnamefont {Bernholc}},\ }\bibfield  {title} {\bibinfo {title} {Theory
  of surface morphology of wurtzite {GaN}(0001) surfaces},\ }\href
  {https://doi.org/10.1103/PhysRevB.56.R12725} {\bibfield  {journal} {\bibinfo
  {journal} {Phys. Rev. B}\ }\textbf {\bibinfo {volume} {56}},\ \bibinfo
  {pages} {R12725} (\bibinfo {year} {1997})}\BibitemShut {NoStop}%
\bibitem [{\citenamefont {Smith}\ \emph {et~al.}(1998)\citenamefont {Smith},
  \citenamefont {Feenstra}, \citenamefont {Greve}, \citenamefont {Shin},
  \citenamefont {Skowronski}, \citenamefont {Neugebauer},\ and\ \citenamefont
  {Northrup}}]{Feenstra1998}%
  \BibitemOpen
  \bibfield  {author} {\bibinfo {author} {\bibfnamefont {A.~R.}\ \bibnamefont
  {Smith}}, \bibinfo {author} {\bibfnamefont {R.~M.}\ \bibnamefont {Feenstra}},
  \bibinfo {author} {\bibfnamefont {D.~W.}\ \bibnamefont {Greve}}, \bibinfo
  {author} {\bibfnamefont {M.-S.}\ \bibnamefont {Shin}}, \bibinfo {author}
  {\bibfnamefont {M.}~\bibnamefont {Skowronski}}, \bibinfo {author}
  {\bibfnamefont {J.}~\bibnamefont {Neugebauer}},\ and\ \bibinfo {author}
  {\bibfnamefont {J.~E.}\ \bibnamefont {Northrup}},\ }\bibfield  {title}
  {\bibinfo {title} {Reconstructions of {GaN}(0001) and (000$\bar{1}$)
  surfaces: {Ga}-rich metallic structures},\ }\href
  {https://doi.org/10.1116/1.590156} {\bibfield  {journal} {\bibinfo  {journal}
  {J. Vac. Sci. Technol. B}\ }\textbf {\bibinfo {volume} {16}},\ \bibinfo
  {pages} {2242} (\bibinfo {year} {1998})}\BibitemShut {NoStop}%
\bibitem [{\citenamefont {Fritsch}\ \emph {et~al.}(1998)\citenamefont
  {Fritsch}, \citenamefont {Sankey}, \citenamefont {Schmidt},\ and\
  \citenamefont {Page}}]{Fritsch1998}%
  \BibitemOpen
  \bibfield  {author} {\bibinfo {author} {\bibfnamefont {J.}~\bibnamefont
  {Fritsch}}, \bibinfo {author} {\bibfnamefont {O.~F.}\ \bibnamefont {Sankey}},
  \bibinfo {author} {\bibfnamefont {K.~E.}\ \bibnamefont {Schmidt}},\ and\
  \bibinfo {author} {\bibfnamefont {J.~B.}\ \bibnamefont {Page}},\ }\bibfield
  {title} {\bibinfo {title} {Ab initio calculation of the stoichiometry and
  structure of the (0001) surfaces of {GaN} and {AlN}},\ }\href
  {https://doi.org/10.1103/PhysRevB.57.15360} {\bibfield  {journal} {\bibinfo
  {journal} {Phys. Rev. B}\ }\textbf {\bibinfo {volume} {57}},\ \bibinfo
  {pages} {15360} (\bibinfo {year} {1998})}\BibitemShut {NoStop}%
\bibitem [{\citenamefont {Audemard}\ and\ \citenamefont
  {Simon}(2009)}]{Glucose}%
  \BibitemOpen
  \bibfield  {author} {\bibinfo {author} {\bibfnamefont {G.}~\bibnamefont
  {Audemard}}\ and\ \bibinfo {author} {\bibfnamefont {L.}~\bibnamefont
  {Simon}},\ }\bibfield  {title} {\bibinfo {title} {Predicting learnt clauses
  quality in modern {SAT} solvers},\ }in\ \href@noop {} {\emph {\bibinfo
  {booktitle} {Proceedings of the 21st International Joint Conference on
  Artificial Intelligence (IJCAI-09)}}},\ \bibinfo {editor} {edited by\
  \bibinfo {editor} {\bibfnamefont {C.}~\bibnamefont {Boutilier}}}\
  (\bibinfo {publisher} {AAAI Press},\ \bibinfo {address} {Menlo Park,
  CA},\ \bibinfo {year} {2009})\ pp.\ \bibinfo {pages}
  {399--404}\BibitemShut {NoStop}%
\bibitem [{\citenamefont {Kuboyama}\ and\ \citenamefont
  {Kusaba}(2025)}]{Kuboyama2024}%
  \BibitemOpen
  \bibfield  {author} {\bibinfo {author} {\bibfnamefont {T.}~\bibnamefont
  {Kuboyama}}\ and\ \bibinfo {author} {\bibfnamefont {A.}~\bibnamefont
  {Kusaba}},\ }\bibfield  {title} {\bibinfo {title} {{SAT} solver-driven
  approach for validating local electron counting rule},\ }\href
  {https://doi.org/10.1016/j.jcrysgro.2024.127927} {\bibfield  {journal}
  {\bibinfo  {journal} {J. Cryst. Growth}\ }\textbf {\bibinfo {volume} {650}},\
  \bibinfo {pages} {127927} (\bibinfo {year} {2025})}\BibitemShut {NoStop}%
\bibitem [{\citenamefont {Northrup}\ and\ \citenamefont
  {Neugebauer}(2004)}]{Northrup2004}%
  \BibitemOpen
  \bibfield  {author} {\bibinfo {author} {\bibfnamefont {J.~E.}\ \bibnamefont
  {Northrup}}\ and\ \bibinfo {author} {\bibfnamefont {J.}~\bibnamefont
  {Neugebauer}},\ }\bibfield  {title} {\bibinfo {title} {Strong affinity of
  hydrogen for the {GaN}(000$\bar{1}$) surface: Implications for molecular beam
  epitaxy and metalorganic chemical vapor deposition},\ }\href
  {https://doi.org/10.1063/1.1808227} {\bibfield  {journal} {\bibinfo
  {journal} {Appl. Phys. Lett.}\ }\textbf {\bibinfo {volume} {85}},\ \bibinfo
  {pages} {3429} (\bibinfo {year} {2004})}\BibitemShut {NoStop}%
\bibitem [{\citenamefont {Hjorth~Larsen}\ \emph {et~al.}(2017)\citenamefont
  {Hjorth~Larsen} \emph {et~al.}}]{ASE2017}%
  \BibitemOpen
  \bibfield  {author} {\bibinfo {author} {\bibfnamefont {A.}~\bibnamefont
  {Hjorth~Larsen}} \emph {et~al.},\ }\bibfield  {title} {\bibinfo {title} {The
  atomic simulation environment---a {Python} library for working with atoms},\
  }\href {https://doi.org/10.1088/1361-648X/aa680e} {\bibfield  {journal}
  {\bibinfo  {journal} {J. Phys.: Condens. Matter}\ }\textbf {\bibinfo {volume}
  {29}},\ \bibinfo {pages} {273002} (\bibinfo {year} {2017})}\BibitemShut
  {NoStop}%
\bibitem [{\citenamefont {Wood}\ \emph {et~al.}(2025)\citenamefont {Wood},
  \citenamefont {Dzamba} \emph {et~al.}}]{UMA2024}%
  \BibitemOpen
  \bibfield  {author} {\bibinfo {author} {\bibfnamefont {B.~M.}\ \bibnamefont
  {Wood}}, \bibinfo {author} {\bibfnamefont {M.}~\bibnamefont {Dzamba}}, \emph
  {et~al.},\ }\bibfield {title} {\bibinfo {title} {{UMA}: A family of
  universal models for atoms},\ }in\ \href
  {https://doi.org/10.52202/085713-4310} {\emph {\bibinfo {booktitle}
  {Advances in Neural Information Processing Systems 38 (NeurIPS 2025)}}}\
  (\bibinfo {year} {2025})\ pp.\ \bibinfo {pages} {129391--129427},\
  \Eprint {https://arxiv.org/abs/2506.23971} {arXiv:2506.23971}
  \BibitemShut {NoStop}%
\bibitem [{\citenamefont {Iwata}\ \emph {et~al.}(2010)\citenamefont {Iwata},
  \citenamefont {Takahashi}, \citenamefont {Oshiyama}, \citenamefont {Boku},
  \citenamefont {Shiraishi}, \citenamefont {Okada},\ and\ \citenamefont
  {Yabana}}]{Iwata2010}%
  \BibitemOpen
  \bibfield  {author} {\bibinfo {author} {\bibfnamefont {J.-I.}\ \bibnamefont
  {Iwata}}, \bibinfo {author} {\bibfnamefont {D.}~\bibnamefont {Takahashi}},
  \bibinfo {author} {\bibfnamefont {A.}~\bibnamefont {Oshiyama}}, \bibinfo
  {author} {\bibfnamefont {T.}~\bibnamefont {Boku}}, \bibinfo {author}
  {\bibfnamefont {K.}~\bibnamefont {Shiraishi}}, \bibinfo {author}
  {\bibfnamefont {S.}~\bibnamefont {Okada}},\ and\ \bibinfo {author}
  {\bibfnamefont {K.}~\bibnamefont {Yabana}},\ }\bibfield  {title} {\bibinfo
  {title} {A massively-parallel electronic-structure calculations based on
  real-space density functional theory},\ }\href
  {https://doi.org/10.1016/j.jcp.2009.11.038} {\bibfield  {journal} {\bibinfo
  {journal} {J. Comput. Phys.}\ }\textbf {\bibinfo {volume} {229}},\ \bibinfo
  {pages} {2339} (\bibinfo {year} {2010})}\BibitemShut {NoStop}%
\bibitem [{\citenamefont {Perdew}\ \emph {et~al.}(1996)\citenamefont {Perdew},
  \citenamefont {Burke},\ and\ \citenamefont {Ernzerhof}}]{PBE1996}%
  \BibitemOpen
  \bibfield  {author} {\bibinfo {author} {\bibfnamefont {J.~P.}\ \bibnamefont
  {Perdew}}, \bibinfo {author} {\bibfnamefont {K.}~\bibnamefont {Burke}},\ and\
  \bibinfo {author} {\bibfnamefont {M.}~\bibnamefont {Ernzerhof}},\ }\bibfield
  {title} {\bibinfo {title} {Generalized gradient approximation made simple},\
  }\href {https://doi.org/10.1103/PhysRevLett.77.3865} {\bibfield  {journal}
  {\bibinfo  {journal} {Phys. Rev. Lett.}\ }\textbf {\bibinfo {volume} {77}},\
  \bibinfo {pages} {3865} (\bibinfo {year} {1996})}\BibitemShut {NoStop}%
\bibitem [{\citenamefont {Soler}\ \emph {et~al.}(2002)\citenamefont {Soler},
  \citenamefont {Artacho}, \citenamefont {Gale}, \citenamefont {Garc\'{i}a},
  \citenamefont {Junquera}, \citenamefont {Ordej\'{o}n},\ and\ \citenamefont
  {S\'{a}nchez-Portal}}]{Siesta2002}%
  \BibitemOpen
  \bibfield  {author} {\bibinfo {author} {\bibfnamefont {J.~M.}\ \bibnamefont
  {Soler}}, \bibinfo {author} {\bibfnamefont {E.}~\bibnamefont {Artacho}},
  \bibinfo {author} {\bibfnamefont {J.~D.}\ \bibnamefont {Gale}}, \bibinfo
  {author} {\bibfnamefont {A.}~\bibnamefont {Garc\'{i}a}}, \bibinfo {author}
  {\bibfnamefont {J.}~\bibnamefont {Junquera}}, \bibinfo {author}
  {\bibfnamefont {P.}~\bibnamefont {Ordej\'{o}n}},\ and\ \bibinfo {author}
  {\bibfnamefont {D.}~\bibnamefont {S\'{a}nchez-Portal}},\ }\bibfield  {title}
  {\bibinfo {title} {The {SIESTA} method for \emph{ab initio} order-{$N$}
  materials simulation},\ }\href {https://doi.org/10.1088/0953-8984/14/11/302}
  {\bibfield  {journal} {\bibinfo  {journal} {J. Phys.: Condens. Matter}\
  }\textbf {\bibinfo {volume} {14}},\ \bibinfo {pages} {2745} (\bibinfo {year}
  {2002})}\BibitemShut {NoStop}%
\bibitem [{\citenamefont {Pedroza}\ \emph {et~al.}(2009)\citenamefont
  {Pedroza}, \citenamefont {da~Silva},\ and\ \citenamefont
  {Capelle}}]{Pedroza2009}%
  \BibitemOpen
  \bibfield  {author} {\bibinfo {author} {\bibfnamefont {L.~S.}\ \bibnamefont
  {Pedroza}}, \bibinfo {author} {\bibfnamefont {A.~J.~R.}\ \bibnamefont
  {da~Silva}},\ and\ \bibinfo {author} {\bibfnamefont {K.}~\bibnamefont
  {Capelle}},\ }\bibfield  {title} {\bibinfo {title} {Gradient-dependent
  density functionals of the {Perdew}-{Burke}-{Ernzerhof} type for atoms,
  molecules, and solids},\ }\href {https://doi.org/10.1103/PhysRevB.79.201106}
  {\bibfield  {journal} {\bibinfo  {journal} {Phys. Rev. B}\ }\textbf {\bibinfo
  {volume} {79}},\ \bibinfo {pages} {201106} (\bibinfo {year}
  {2009})}\BibitemShut {NoStop}%
\bibitem [{\citenamefont {Odashima}\ \emph {et~al.}(2009)\citenamefont
  {Odashima}, \citenamefont {Capelle},\ and\ \citenamefont
  {Trickey}}]{Odashima2009}%
  \BibitemOpen
  \bibfield  {author} {\bibinfo {author} {\bibfnamefont {M.~M.}\ \bibnamefont
  {Odashima}}, \bibinfo {author} {\bibfnamefont {K.}~\bibnamefont {Capelle}},\
  and\ \bibinfo {author} {\bibfnamefont {S.~B.}\ \bibnamefont {Trickey}},\
  }\bibfield  {title} {\bibinfo {title} {Tightened {Lieb}-{Oxford} bound for
  systems of fixed particle number},\ }\href
  {https://doi.org/10.1021/ct8005634} {\bibfield  {journal} {\bibinfo
  {journal} {J. Chem. Theory Comput.}\ }\textbf {\bibinfo {volume} {5}},\
  \bibinfo {pages} {798} (\bibinfo {year} {2009})}\BibitemShut {NoStop}%
\bibitem [{\citenamefont {Duzik}\ \emph {et~al.}(2013)\citenamefont {Duzik},
  \citenamefont {Thomas}, \citenamefont {Van~der Ven},\ and\ \citenamefont
  {Millunchick}}]{DuzikMillunchick2013}%
  \BibitemOpen
  \bibfield  {author} {\bibinfo {author} {\bibfnamefont {A.}~\bibnamefont
  {Duzik}}, \bibinfo {author} {\bibfnamefont {J.~C.}\ \bibnamefont {Thomas}},
  \bibinfo {author} {\bibfnamefont {A.}~\bibnamefont {Van~der Ven}},\ and\
  \bibinfo {author} {\bibfnamefont {J.~M.}\ \bibnamefont {Millunchick}},\
  }\bibfield  {title} {\bibinfo {title} {Surface reconstruction stability and
  configurational disorder on {Bi}-terminated {GaAs}(001)},\ }\href
  {https://doi.org/10.1103/PhysRevB.87.035313} {\bibfield  {journal} {\bibinfo
  {journal} {Phys. Rev. B}\ }\textbf {\bibinfo {volume} {87}},\ \bibinfo
  {pages} {035313} (\bibinfo {year} {2013})}\BibitemShut {NoStop}%
\bibitem [{\citenamefont {Kusaba}(2025)}]{PyAPX}%
  \BibitemOpen
  \bibfield  {author} {\bibinfo {author} {\bibfnamefont {A.}~\bibnamefont
  {Kusaba}},\ }\href@noop {} {\bibinfo {title} {{PyAPX}: {Python} toolkit for
  atomic configuration pattern exploration}},\ \bibinfo {howpublished}
  {\url{https://github.com/a-ksb/PyAPX}} (\bibinfo {year} {2025})\BibitemShut
  {NoStop}%
\bibitem [{\citenamefont {Kawahara}\ \emph {et~al.}(2017)\citenamefont
  {Kawahara}, \citenamefont {Inoue}, \citenamefont {Iwashita},\ and\
  \citenamefont {Minato}}]{Kawahara2017}%
  \BibitemOpen
  \bibfield  {author} {\bibinfo {author} {\bibfnamefont {J.}~\bibnamefont
  {Kawahara}}, \bibinfo {author} {\bibfnamefont {T.}~\bibnamefont {Inoue}},
  \bibinfo {author} {\bibfnamefont {H.}~\bibnamefont {Iwashita}},\ and\
  \bibinfo {author} {\bibfnamefont {S.}~\bibnamefont {Minato}},\ }\bibfield
  {title} {\bibinfo {title} {Frontier-based search for enumerating all
  constrained subgraphs with compressed representation},\ }\href
  {https://doi.org/10.1587/transfun.E100.A.1773} {\bibfield  {journal}
  {\bibinfo  {journal} {IEICE Trans. Fundamentals}\ }\textbf {\bibinfo {volume}
  {E100.A}},\ \bibinfo {pages} {1773} (\bibinfo {year} {2017})}\BibitemShut
  {NoStop}%
\bibitem [{\citenamefont {McKay}(1998)}]{McKay1998}%
  \BibitemOpen
  \bibfield  {author} {\bibinfo {author} {\bibfnamefont {B.~D.}\ \bibnamefont
  {McKay}},\ }\bibfield  {title} {\bibinfo {title} {Isomorph-free exhaustive
  generation},\ }\href {https://doi.org/10.1006/jagm.1997.0898} {\bibfield
  {journal} {\bibinfo  {journal} {J. Algorithms}\ }\textbf {\bibinfo {volume}
  {26}},\ \bibinfo {pages} {306} (\bibinfo {year} {1998})}\BibitemShut
  {NoStop}%
\bibitem [{\citenamefont {Dreyer}\ \emph {et~al.}(2014)\citenamefont {Dreyer},
  \citenamefont {Janotti},\ and\ \citenamefont {Van~de Walle}}]{Dreyer2014}%
  \BibitemOpen
  \bibfield  {author} {\bibinfo {author} {\bibfnamefont {C.~E.}\ \bibnamefont
  {Dreyer}}, \bibinfo {author} {\bibfnamefont {A.}~\bibnamefont {Janotti}},\
  and\ \bibinfo {author} {\bibfnamefont {C.~G.}\ \bibnamefont {Van~de Walle}},\
  }\bibfield  {title} {\bibinfo {title} {Absolute surface energies of polar and
  nonpolar planes of {GaN}},\ }\href
  {https://doi.org/10.1103/PhysRevB.89.081305} {\bibfield  {journal} {\bibinfo
  {journal} {Phys. Rev. B}\ }\textbf {\bibinfo {volume} {89}},\ \bibinfo
  {pages} {081305(R)} (\bibinfo {year} {2014})}\BibitemShut {NoStop}%
\bibitem [{\citenamefont {Bermudez}(2017)}]{Bermudez2017}%
  \BibitemOpen
  \bibfield  {author} {\bibinfo {author} {\bibfnamefont {V.~M.}\ \bibnamefont
  {Bermudez}},\ }\bibfield  {title} {\bibinfo {title} {The fundamental surface
  science of wurtzite gallium nitride},\ }\href
  {https://doi.org/10.1016/j.surfrep.2017.05.001} {\bibfield  {journal}
  {\bibinfo  {journal} {Surf. Sci. Rep.}\ }\textbf {\bibinfo {volume} {72}},\
  \bibinfo {pages} {147} (\bibinfo {year} {2017})}\BibitemShut {NoStop}%
\bibitem [{\citenamefont {Kuboyama}\ \emph {et~al.}(2026)\citenamefont
  {Kuboyama}, \citenamefont {Kusaba}, \citenamefont {Kawka},\ and\
  \citenamefont {Kempisty}}]{GaNTilingData2026}%
  \BibitemOpen
  \bibfield  {author} {\bibinfo {author} {\bibfnamefont {T.}~\bibnamefont
  {Kuboyama}}, \bibinfo {author} {\bibfnamefont {A.}~\bibnamefont {Kusaba}},
  \bibinfo {author} {\bibfnamefont {K.}~\bibnamefont {Kawka}},\ and\ \bibinfo
  {author} {\bibfnamefont {P.}~\bibnamefont {Kempisty}},\ }\href@noop {}
  {\bibinfo {title} {Data and code for: {Tiling} decomposition multiplicity
  predicts stability of {GaN}(0001) surface reconstructions}},\ \bibinfo
  {note} {version v1.0.0}\ (\bibinfo {publisher} {Zenodo},\ \bibinfo
  {year} {2026}),\ \href {https://doi.org/10.5281/zenodo.24357550}
  {doi:10.5281/zenodo.24357550};\ development repository at
  \url{https://github.com/tkub/2026gan-tiling}\BibitemShut {NoStop}%
\end{thebibliography}
\end{document}